\documentclass[12pt]{article}
\usepackage[dvips]{graphicx}
\usepackage{amssymb}
\usepackage{amsmath}

\def\beq{\begin{equation}}\def\eeq{\end{equation}}
\def\bea{\begin{eqnarray}}\def\eea{\end{eqnarray}}

\begin{document}

\title{Electromagnetic Lagrangian on a causal set that resides on edges rather than points}
\author{Roman Sverdlov,
\\Department of Mathematics, University of New Mexico} 

\date{May 23, 2018}
\maketitle

\begin{abstract}
The goal of this paper is to introduce one of the versions of the electromagnetic Lagrangian on a causal set in such a way that would address the non-locality issues inherent to causal set theory. The key idea is that Lagrangian density is assigned to the edges rather than points, and there is a way of defining the concept of ``neighboring edges" of a given edge in such a way that each edge has only finitely many neighboring edges which would ultimately allow for the theory to be local. That is to be contrasted with points where every point has infinitely many direct neighbors which is a source of non-locality. The edges are needed in order to define electromagnetic Lagrangian anyway, regardless of the consideration of locality; the novelty of this paper is to assign Lagrangian density to the edges as well. Also, in the other papers edges were both spacelike and timelike, while in this paper they are only timelike. This makes calculations considerably more complicated, but it is crucial in preserving locality since the Lorentz group in a hyperplane perpendicular to the edge is compact only if the edge is timelike. 
\end{abstract}

\subsection*{Introduction}

Causal set theory, originally proposed by Rafael Sorkin, is a discrete model of spacetime in which the only geometry consists of the set of events and causal relations (for more detailed review, see \cite{Review1}, \cite{Review2}, \cite{Review3}, \cite{Review4}) We say that $x \prec y$ if we can go from $x$ to $y$ without going faster than the speed of light, and then we drop the word ``speed" and just think of $\prec$ as partial ordering on a general set. It has been shown by Malement and Hawking (see \cite{Malament} and \cite{Hawking}) that if we know the causal relations between all of the points on a spacetime manifold, we can deduce the metric, up to Weyl scaling. In discrete case, the Weyl scaling can be deduced from the count of the number of points, under the assumption that every point takes up the same volume. Thus, a complete information about the discretized geometry is obtained. At the same time, on the small scales the geometry as such might break down due to quantum fluctuations, yet one might hypothesize that the partial order would persist even then; if so, this would make partial order more fundamental. 

However, due to the fact that the partial ordering is the only fundamental geometry available, we are not allowed to use Lorentz indexes. This means that all of the physical entities should be rewritten. For example, Gauge field $A^{\mu}$ should be replaced with a two-point function 
\beq a(p,q) = \int_{\gamma (p,q)} g_{\mu \nu} A^{\mu} dx^{\nu} \eeq
where $\gamma (p,q)$ is a geodesic connecting $p$ and $q$. The expression on the right hand side, of course, is ill defined due to the presence of Lorentzian indexes; thus it should be eventually dropped in favor of the left hand side that doesn't include them. Similarly, any expression containing derivative signs should be replaced with some other expression that doesn't have them. Some examples of this type of work can be found here: \cite{Dambertian1},  \cite{Dambertian2}, \cite{GravityAndMatter}, \cite{Gauge}, \cite{Bosonic}. 

One common problem that the above cited papers face is the one of non-locality. This is due to the fact that the Lorentzian neighborhood of a point is the vicinity of its lightcone, which has infinite volume. If we simulate causal set via Poisson distribution of points on a manifold, we would expect with absolute certainty that any given point will have infinitely many direct neighbors that are separated arbitrarily far away coordinate-wise. In order to avoid it, the distribution of points shouldn't be Poisson; but in this case it would single out some sort of preferred frame. Thus, there is a trade-off between wanting to avoid preferred frame and wanting to avoid non-locality; in fact, a theorem was proven that one can't avoid both things at the same time (see \cite{NonlocalityTheorem}): in particular, if the number of neighbors is finite, then one of them would be nearest, which would lead to preferred choice of the coordinate axis (namely, the one that has the first point at the origin, and passes through its nearest neighbor), while if the number of neighbors is infinite, then we would have nonlocality. 

Of course, one could have argued, as was done in \cite{Smolin1} and \cite{Smolin2}, that if spacetime is emergent rather than fundamental, then the above questions are moot: on microscopic scale we simply don't have such a thing as lightcone, just like on sub-molecular scale we don't have such a thing as smell. This, however, is not the rout that causal set theory took: even though, according to causal set theory, coordinates aren't viewed as fundamental, they do, very much, aim at the special case of a causal set being simulated by Poisson scatter of points on a manifold and, indeed, making this special case work is one of the main goals of causal set theory. 

This being the case, the past work on causal set Lagrangians have taken two main approaches. One was to openly admit non-locality and hope that near-lightcone contributions would cancel (\cite{Dambertian1} and \cite{Dambertian2}), the other was to suggest that ``preferred frames" are dynamically-selected based on the specific behavior of the fields and, therefore, those frames aren't preferred since they would alternate as our fields change in our path integral (\cite{GravityAndMatter}, \cite{Gauge}, \cite{Bosonic}). The former approach is, indeed, linear, but the assertion that non-localities truly cancel can be questioned; the second approach is more local, but it has non-linear effects that would, essentially, preclude the possibility of taking path integral analytically, and would limit us to a numeric simulations on a toy model with limited number of points. Nevertheless, some work has been done by Johnston in defining propagators without the need of Lagrangians on the first place (see \cite{Johnston1} and \cite{Johnston2}) and, indeed, that approach has successfully avoided any type of non-locality in any form, although it might be a bit more difficult (albeit not impossible) to generalize it to gauge fields. 

In any case, as far as the current paper is concerned, we are proposing to use Lagrangian approach, but address the issue of non-locality in a different way than what was done before. In particular, we propose that the Lagrangian density doesn't reside at a point but, instead, it resides on the timelike edge between the two points (and then the action will be the sum over all edges); similarly, scalar field will reside on edges as well. Here, by an edge, we mean a pair of points $p \prec^* q$ where $\prec^*$ is defined in the following way: $p \prec^* q$ is true if and only if $p \prec q$ is true and there is no $r$ satisfying $p \prec r \prec q$. We introduce a notion of a ``neighbor" of an edge: namely, an edge $b \prec^* c$ is a ``neighbor" of an edge $a \prec^* b$ if the number of points $d$ satisfying $a \prec d \prec c$ is below some upper bound. Based on this definition, two edges can share a point and still not be neighbors of each other. Consequently, any edge will be shown to have only finitely many edge-neighbors, in contrast to a point that has infinitely many point neighbors. Thus, by placing all the physical information (including Lagrangian density, both scalar and vector fields, etc) on edges rather than points we restore locality. At the same time, the points are still relevant since we can put sources/sinks at those points as opposed to edges. In a toy model with just the scalar field $\phi$, our action will look like this: 
\beq S (\phi, J) = \sum_{p \prec^* q} ({\cal L} (p,q) + (J(p)+J(q)) \phi (p,q)) \eeq
leading to the partition function 
\beq Z (J)= \int [{\cal D} \phi] e^{i S(\phi, J)} \eeq
where 
\beq [{\cal D} \phi] = \prod_{p \prec^* q} d \phi (p, q) \eeq
leading to a propagator 
\beq D(r,s) = \frac{\partial^2}{\partial J(r) \partial J(s)}\bigg\vert_{\forall x (J(x)=0)}  \ln Z(J) \eeq
Since those sources/sinks aren't directly coupled to each other, the non-locality of the points doesn't stand in our way of performing a calculation of a propagator. 

On a continuum limit, the direction of an edge will correspond to the tangent vector $v$, which means that Lagrangian density will take a form of ${\cal L} (\phi, a; x, v)$ as opposed to ${\cal L} (\phi, a; x)$, where $x$ is a location in spacetime. In and of itself, ${\cal L} (\phi, a; x, v)$ doesn't violate Lorentz invariance any more than ${\cal L} (\phi, a; x)$ violates the translational invariance, since in the process of going from $\cal L$ to $S$ we will be integrating over $v$, just like we will be integrating over $x$. However, since the Lagrangians that we know happen to take the form of ${\cal L} (\phi, a; x)$, we have to say that the $v$-dependence is very small. This can be accomplished by arranging so that it cancels. 

Let us go ahead and describe the Lagrangian that would accomplish such a cancellation:  Let set $C_1$ be a set of contours $prqsp$ such that $p \prec r \prec q$, $p \prec s \prec q$, the relation between $r$ and $s$ is not specified, and there are no points $x$ satisfying $p \prec x \prec q$ other than $r$ and $s$. Furthermore, let set $C_2$ consist of the set of contours $prsqp$ such that $p \prec r \prec s \prec q \prec p$ such that there is no point $x$ satisfying $p \prec x \prec q$ other than $r$ and $s$ (henceforth, for the sake of brevity, we will denote the contours of $C_1$ as $p \prec r \prec q \succ s \succ p$ and contours of $C_2$ as $p \prec r \prec s \prec q \succ p$, where $x \succ y$ is the same as $y \prec x$). We are looking for the expression of the form 
\beq S = A \sum_{(prqsp) \in C_1} \Phi^2 (prqsp) + B \sum_{(prsqp) \in C_2} \Phi^2 (prsqp) \label{IntroCombination} \eeq
where
\beq \Phi (x_1, \cdots, x_n, x_1) = a (x_1, x_2) +  \cdots + a (x_{n-1}, x_n) + a (x_n, x_1) \label{FluxIntro} \eeq
is a \emph{flux} through the contour $(x_1, \cdots, x_n, x_1)$ which, again, coincides with the flux as we know it (the surface integral) in the manifold case yet remains well defined in non-manifold case (where surface integral isn't defined to begin with). If we take a single contour, clearly, we can't expect the above expression to approximate anything since the location of points is random. But if we take a collection of contours $\{(prqsp) \in C_1 \vert (q^{\mu} - p^{\mu})v_{\mu}=0 \}$ and  $\{(prsqp) \in C_2 \vert (q^{\mu} - p^{\mu})v_{\mu}=0 \}$, and assume that the field is approximately linear in the region where we took that collection of points (which is really the consequence of the fact that it is differentiable and the region we took is small), then we would statistically expect to have an expression of the form $C F_{00}^2 + D F_{ij} F_{ij}$,  in the reference frame in which $t$-axis is parallel to $v^{\mu}$, where $C$ and $D$ are functions of $A$ and $B$; and then we can adjust $A/B$ in such a way as to get $D/C=-1$ (and as far as actual values of those coefficients rather than rations, that part is a lot less important). 

The adjustment of $A/B$ looks very artificial, which might raise a question as to why doesn't the principle of relativity hold naturally, without artificial adjustment. To answer this question, we can once again point out the fact that if we agree that the presence of $x$ in $\cal L$ doesn't violate translational invariance, we can also agree that the presence of $v$ doesn't violate Lorentz invariance, since we are integrating over both. So by getting rid of $v$-dependence we aren't trying to save Lorentz invariance but, rather, we are trying to adjust our results to the experiment, which didn't detect any $v$-dependence. As a matter of fact, the residual $v$-dependence still remains, even after the cancellation. In particular, if $v$ is very close to the speed of light, we can no longer assume the linearity that is important for above calculation to hold. Intuitively, we can argue that if two spaceships move very close to the speed of light relative to each other, then whatever physical object one spaceship sees is invisible to the other spaceship due to the Lorentz contraction that takes a mile to a micron, the latter being invisible. We have to assume that invisibility, however, since we are assuming that whatever ``is" visible is sufficiently smooth for our calculations to hold. This implies the type of $v$ dependence that can't be read off from Lorentz transformations. We are okay with this $v$-dependence relativity-wise, due to the analogy with $x$-dependence that we talked about, so the only question is whether or not it can be experimentally falsified, and it can't, since we can't reach such high velocities. 

It should be noticed that analogous (with much simpler calculations) adjustments of coefficients have been done in \cite{GravityAndMatter}, \cite{Gauge} and \cite{Bosonic}. The key difference between what we are doing here and what was done there is that in those papers the Lagrangian was at a point, and the preferred frame was a function of the fields; on the other hand, in a current paper the Lagrangian is at an edge and, therefore, preferred frame is independent of the behavior of the fields (it coincides with the direction of the edge, regardless of what the fields do). In case of \cite{GravityAndMatter}, \cite{Gauge} and \cite{Bosonic} the Lorentz invariance is restored in the step of integration over the fields (once we already have the action) while according to the current paper it is restored at the earlier step, when we integrate over space-velocity to get an action. Consequently, the nonlinear effects in \cite{GravityAndMatter}, \cite{Gauge} and \cite{Bosonic} have been avoided in this paper. 

The other difference between this paper and the ones I just mentioned is that in \cite{Gauge} and \cite{Bosonic} the holonomy corresponding to the gauge field, $a(r,s)$, was defined regardless of whether $r$ and $s$ are causally related; on the other hand, in this paper $a(r,s)$ is only defined when they are, in fact, causally related. This is the main reason the calculations in this paper are considerably more complicated, but I consider it worth it since the assumption that $r$ and $s$ are causally related makes $a(r,s)$ more physically meaningful. Apart from that, spacelike edge doesn't avoid the issue of non-locality since the Lorentz group on a hyperplane perpendicular to spacelike edge isn't compact; in case of timelike edge it is.

One might wonder that, due to the ``nearest neighbor arguments" in \cite{NonlocalityTheorem}, the edges, being local, would violate relativity. First of all, it is important to point out that an edge inherently has a preferred frame, namely its own direction, and it doesn't violate relativity any more than the location of a point violates translational invariance. However, the argument in \cite{NonlocalityTheorem} translates into a different problem: the discrete effects would result in a ``preferred accelerations" (other than zero) associated with various velocities. In particular, the direction of the edge would correspond to velocity, and the way it compares to the direction of its nearest edge-neighbor would correspond to the ``preferred acceleration". This type of concern, however, can be addressed in the following way: one can simply point out that any other version of causal set theory would also contain both points and edges, and the edges would always play ``some" role -- at the very least, in defining gauge field. The only thing we have changed is exactly ``how much" role do edges play. From this perspective, whatever violation of relativity that we might have due to edges in a given paper, would also apply to any other version of causal set theory, which, on a flip side, means that we can claim to satisfy the relativity standard used in those other papers. That ``relativity standard" is satisfied by the presence of points, which don't have nearest neighbors. 

Speaking of ``standards", it is important to note that, regardless of any of those issues, the Lorentz invariance can't possibly be accomplished in the strictest sense, since a discrete space can't be mapped onto itself via Lorentz transformation. Thus, we can simply use \cite{NonlocalityTheorem} as a guideline of the ``imperfect" standard to aim for. According to that paper, Poisson scatter on Minkowski space is Lorentz invariant, while Poisson scatter in Euclidean space isn't. In both Minkowski and Eucledian spaces the edges have nearest neighbors. Furthermore, in Minkowski space the points have hierarchy of neighbors, some nearer than others, they just don't have ``the" nearest ones. In Euclidean space the points (as opposed to edges) have ``the" nearest neighbors (as opposed to hierarchy of neighbors). The latter is apparently the feature in \cite{NonlocalityTheorem} that they want to avoid, and it was successfully avoided in this paper.

\subsection*{Locality of the edges: more saddle issues}

Before we go to the main part of the paper and do the calculation just described, let us address some more subtle non-locality issue: do the edges, indeed, have finitely many edge neighbors? In light of the fluctuation of Poisson distribution, the answer to this question might be more difficult than it might first seem.  We would like to be able to say that if $a \prec b \prec c$ then the edge $a \prec b$ isn't a neighbor of the edge $b \prec c$ if $c$ is too far away from $a$. But, since we aren't able to measure the distance independently of counting points, what we might say is that the number of points in \emph{Alexandrov set} 
\beq \alpha (a,c) = \{r \vert a \prec r \prec c \} \eeq
has to be less than or equal to $n$ in order for the above two edges to be neighbors. But in light of Poisson distribution it is possible that \emph{despite} the fact that the distance between $a$ and $c$ is large, it just happened ``by accident" that none of the randomly scattered points happened to fall in $\alpha (a, c)$. And, as a result, we would end up ``mistakenly" counting the edge $(b,c)$ as a neighbor of an edge $(a, b)$, despite the fact that we shouldn't have. Sure, such ``accident" is highly unlikely. But since the volume of space is infinite, we have infinitely many ``opportunities" for such an ``accident", which makes logically possible for the expectation value of the number of such unlikely events to be infinite as well. Yet, the region of space out of which the ``true neighbors" are selected is finite, so the expectation number of ``true neighbors" is finite which is obviously infinitely smaller than the expectation number of ``false neighbors". Luckily, this problem is avoided because, as it turns out, the above described ``unlikely events" aren't ``equally unlikely" but, instead, they are progressively ``less and less" likely, which means that their expected number is a convergent series. Let us, therefore, do a calculation that would show that this is the case. 

Let us fix points $a$ and $b$, and look at all possible points $c$ to see how many edges-neighbors ``bc" does the edge ``ab" statistically is expected to have. On the surface, it might seem like a difficult problem, because if we have two points $c_1$ and $c_2$ and $\alpha (a,c_1)$ overlaps with $\alpha (a, c_2)$ then the question whether or not the edge $bc_1$ became a neighbor of $ab$ or the edge $bc_2$ became a neighbor aren't statistically independent. The way to avoid this problem is to point out that if we are talking about expectation number of positive outcomes (as opposed to a probability of at least one positive outcome taking place) then the expectation numbers add, regardless of whether the two events are statistically dependent or not. Suppose the probability of the edge $bc$ becoming a neighbor of $ab$ \emph{if} the point $c$ were to land at $x$ is $\pi(x)$, and the probability that point $c$ lands into the volume $dv$ around $x$ is $\rho (x) d^d x$, then the joint probability of those two things happening  probability density of scatter is $\rho (x) \pi (x) d^d x$; due to the smallness of $d^d x$, the expectation number of such events is $\rho (x) \pi (x) d^d x + 0 ((d^d x)^2)$; the $0 ((d^d x)^2)$ integrates to zero, so the expectation number of neighbor edges becomes 
\beq \langle \rm number \; of \; edge \; neighbors \rangle = \int_{x \succ b} \rho (x) \pi (x) d^d x \eeq
in causal set theory it is assumed that $\rho = const$, so we have 
\beq \langle \rm number \; of \; edge \; neighbors \rangle = \rho \int_{c \succ b}  \pi (x) d^d x \eeq
Now, if the criteria of the edge $bc$ qualifying to be a neighbor of the edge $ab$ is that the number of points in $\alpha (a,c)$ being less than or equal to $n$, then 
\beq \pi_n (c) = e^{- \rho V (a,c)} \sum_{k=0}^n \frac{(\rho V(a,c))^k}{k!} \eeq
where $\rho$ is the density of Poisson distribution and $V(a,c)$ is a volume of $\alpha (a,c)$. Now, we know that 
\beq V(a, c) = k_d \tau^d (a,c) \eeq
for some $k_d$, where $\tau (a,c)$ is the Lorentzian distance from $a$ to $c$, and $d$ is the dimension of the spacetime. Pick a reference frame where $t$-axis passes through $a$ and $b$, and the origin is at $a$. In this frame, 
\beq \tau^2 (a,c) = (t(c)- t(a))^2 - \vert \vec{r} (c) \vert^2 = (t(b) - t(a) + t(c)- t(b))^2 - \vert \vec{r} (c) \vert^2 = \nonumber \eeq
\beq =  (t(b)-t(a))^2 + 2 (t(b)-t(a))(t(c)-t(b))+ (t(c)-t(b))^2 - \vert \vec{r} (c) \vert^2 \eeq 
Since we know that $b \prec c$, we know that 
\beq (t(c)- t(b))^2 - \vert \vec{r} (c) \vert^2 \geq 0 \eeq
and, therefore, 
\beq \tau^2 (a,c) \geq (t(b)-t(a))^2 + 2(t(b)-t(a))(t(c)-t(b)) \eeq
Now, for the volume we need $\tau^d (a,c)$. For that, we write 
\beq \tau^d (a,c) = (t(b)-t(a))^d \bigg(\frac{\tau(a,c)}{t(b)-t(a)} \bigg)^d \geq (t(b)-t(a))^d \bigg(\frac{\tau(a,c)}{t(b)-t(a)} \bigg)^2 = \nonumber \eeq
\beq = (t(b)-t(a))^d \bigg(1+ 2 \frac{t(c)-t(b)}{t(b)-t(a)} \bigg) \eeq 
Now, the volume of $\alpha (a,c)$ is given by 
\beq V (p,q) =k_d \tau^d (p,q) \eeq
for some dimension-based coefficient $k_d$. By substituting the expression for $\tau^d$ we obtain 
\beq V (p,q) = k_d (t(b)-t(a))^d \bigg(1+ 2 \frac{t(c)-t(b)}{t(b)-t(a)} \bigg) \eeq
and, therefore, the expectation value of the number of edges is 
\beq \langle \rm number \; of \; edge \; neighbors \rangle = \nonumber \eeq
\beq = \rho \int_{x \succ b} d^d x \sum_{k=0}^n \frac{(\rho V(a,c))^k}{k!}  \exp \bigg( - \rho k_d (t(b)-t(a))^d \bigg(1+ 2 \frac{t(c)-t(b)}{t(b)-t(a)} \bigg) \bigg) \eeq
Obviously, attempting to evaluate this integral might be very difficult, and there is no need for it. What is clear, however, is that this integral is convergent. This implies that if we find large enough region, we can say that ``most" of the neighbor edges will be found in that region. Since that region is finite,  we can rescale our thinking to see it as ``small", and thus claiming our theory is local. 

\subsection*{Determining $A_{\mu}$ from $F_{\mu \nu}$} 

Up until now we were trying to address some of the conceptual problems related to locality and linearity. Let us now switch gears and, assuming that we agree to use the linearity assumption, and also that we agree to seek the equation of the form Eq.\ref{IntroCombination}, actually try and perform the calculation for $A/B$. As a first step to doing that, we need to find out the expression for $a(p,q)$ given that we assume that $F_{\mu \nu}$ is constant (which would then allow us to express $\Phi$ in terms of $F_{\mu \nu}$ rather than $a$).  The focus of this section is to find an expression of $A^{\mu}$ in a particular gauge, if we know $F_{\mu \nu}$, and the focus of next section will be to find $a (p,q)$ by using $A^{\mu}$ we will have found in this section.

So, let us proceed with finding out the expression for $A^{\mu}$ in terms of $F_{\mu \nu}$. Let us assume that $F_{\mu \nu}$ is constant. We are trying to find the linear expression from $A_{\mu}$. Our guess is that the expression we are looking for takes the form
\beq A_{\mu} = A_{0 \mu} + E_{\mu \nu} x^{\nu} \eeq
Note that $E_{\mu \nu}$ doesn't have to be either symmetric or antisymmetric (which is why we used $E_{\mu \nu}$ instead of $F_{\mu \nu}$). We will use Gauge transformation (Eq \ref{Gauge} onward) in order to absorb the symmetric part of $E_{\mu \nu}$ into the gauge, extract antisymmetric part, and find the coefficient that relates that antisymmetric part to $F_{\mu \nu}$. Let us first compute $F_{\mu \nu}$ for general $E_{\mu \nu}$:
\beq \partial_{\mu} A_{\nu} = \partial_{\mu} (E_{\nu \rho} x^{\rho}) = E_{\nu \rho} \partial_{\mu} x^{\rho} = E_{\nu \rho} \delta^{\rho}{\mu} = E_{\nu \mu} \eeq
By rearranging $\mu$ and $\nu$, we have 
\beq \partial_{\nu} A_{\mu} = E_{\mu \nu} \eeq
and, therefore, 
\beq F_{\mu \nu} = \partial_{\mu} A_{\nu} - \partial_{\nu} A_{\mu} = E_{\nu \mu} - E_{\mu \nu} \label{FfromE} \eeq
Now let us do gauge transformation
\beq A^{\prime}_{\mu} = A_{\mu} + \partial_{\mu} \lambda \label{Gauge} \eeq
Let 
\beq \Lambda = \lambda - c \; E_{\mu \nu} x^{\mu} x^{\nu}  \eeq
and, therefore, 
\beq \lambda= \Lambda + c \; E_{\mu \nu} x^{\mu} x^{\nu} \label{Lambda} \eeq
Then 
\beq \partial_{\mu} \lambda = \partial_{\mu} \Lambda + \partial_{\mu} \big(c \; E_{\rho \sigma} x^{\rho} x^{\sigma} \big) = \partial_{\mu} \Lambda + c \big( E_{\rho \sigma} (\partial_{\mu} x^{\rho}) x^{\sigma} + E_{\rho \sigma} x^{\rho} \partial_{\mu} x^{\sigma} \big) = \nonumber \eeq 
\beq = \partial_{\mu} \Lambda + c \big(E_{\rho \sigma} \delta^{\rho}_{\mu} x^{\sigma} + E_{\rho \sigma} x^{\rho} \delta^{\sigma}_{\mu} \big) = \partial_{\mu} \Lambda + c \big(E_{\mu \sigma} x^{\sigma} + E_{\rho \mu} x^{\rho} \big) = \partial_{\mu} \Lambda + c (E_{\mu \nu} + E_{\nu \mu}) x^{\nu} \eeq
Therefore,
\beq A^{\prime}_{\mu} = A_{\mu} + \partial_{\mu} \lambda = A_{0 \mu} + E_{\mu \nu} x^{\nu} + c (E_{\mu \nu} + E_{\nu \mu}) x^{\nu} + \partial_{\mu} \Lambda = A_{0 \mu} + (1+c)E_{\mu \nu}x^{\nu} + cE_{\nu \mu} x^{\nu} + \partial_{\mu} \Lambda \eeq 
By setting $c=-1/2$ we have 
\beq c=-\frac{1}{2} \Longrightarrow \nonumber \eeq
\beq \Longrightarrow A^{\prime}_{\mu} = A_{0 \mu} + \bigg(1- \frac{1}{2} \bigg) E_{\mu \nu} x^{\nu} - \frac{1}{2} E_{\nu \mu} x^{\nu} + \partial_{\mu} \Lambda = A_{0 \mu} + \frac{1}{2} E_{\mu \nu} x^{\nu} - \frac{1}{2} E_{\nu \mu} x^{\nu} + \partial_{\mu} \Lambda = \nonumber \eeq
\beq = A_{0 \mu} - \frac{1}{2} (E_{\nu \mu} - E_{\mu \nu})x^{\nu} + \partial_{\mu} \Lambda =^{Eq \; \ref{FfromE}} A_{0 \mu} - \frac{1}{2} F_{\mu \nu} x^{\nu} + \partial_{\mu} \Lambda \label{APrimeFromf} \eeq 
From now on we will drop prime and write 
\beq A_{\mu} = A_{0 \mu} - \frac{1}{2} F_{\mu \nu} x^{\nu} + \partial_{\mu} \Lambda \label{AFromF}\eeq 

\subsection*{Equation for holonomy and flux}

Now that we have computed $A^{\mu}$, lets compute holonomy $a(r,s)$, which is given by
\beq a (r,s) = \int_{\gamma (r,s)} g^{\mu \nu} A_{\mu} dx_{\nu} \eeq
where $\gamma (r,s)$ is a geodesic connecting $r$ and $s$. Since we assume space is flat, it is straight line. Lets parametrize that straight line so that $-1$ maps to $r$ and $1$ maps to $s$; thus,  
\beq x(t) = \frac{r+s}{2} + t \frac{s-r}{2} \eeq
and, by substituting it into Eq \ref{AFromF}, we obtain
\beq A_{\mu}= A_{0 \mu} - \frac{1}{2} F_{\mu \nu} x^{\nu} + \partial_{\mu} \Lambda= A_{0 \mu} - \frac{1}{2} F_{\mu \nu} \bigg(\frac{r^{\nu}+s^{\nu}}{2} + t \frac{s^{\nu}-r^{\nu}}{2} \bigg) + \partial_{\mu} \Lambda \eeq 
Apart from that, we also know that 
\beq dx = \frac{s-r}{2} dt \eeq
By making those substitutions, we obtain
\beq a(r,s) = \int_{\gamma (r,s)} \partial_{\mu} \Lambda \; dx^{\mu} +  \int_{\gamma (r,s)} g^{\mu \nu} \bigg(A_{0 \mu} - \frac{1}{2} F_{\mu \rho} \bigg(\frac{r^{\rho}+s^{\rho}}{2} + t \frac{s^{\rho}-r^{\rho}}{2} \bigg)  \bigg) \bigg(\frac{s_{\nu}-r_{\nu}}{2} dt \bigg) = \nonumber \eeq
\beq =  \Lambda (s) - \Lambda (r) + g^{\mu \nu} \bigg(A_{0 \mu} \int_{-1}^1 dt - \frac{1}{2} F_{\mu \rho} \bigg(\frac{r^{\rho}+s^{\rho}}{2} \int_{-1}^1 dt +  \frac{s^{\rho}-r^{\rho}}{2} \int_{-1}^1 tdt \bigg) \bigg) \bigg(\frac{s_{\nu}-r_{\nu}}{2} \bigg)  = \nonumber \eeq
\beq =  \Lambda (s) - \Lambda (r) + g^{\mu \nu} \bigg(2A_{0 \mu}- \frac{1}{2} F_{\mu \rho} \bigg(\frac{r^{\rho}+s^{\rho}}{2} 2 +  \frac{s^{\rho}-r^{\rho}}{2} 0  \bigg) \bigg) \bigg(\frac{s_{\nu}-r_{\nu}}{2} \bigg)  = \nonumber \eeq
\beq = \Lambda (s) - \Lambda (r) +g^{\mu \nu} A_{0 \mu} (s_{\nu} - r_{\nu})- \frac{1}{4} g^{\mu \nu} F_{\mu \rho} (r^{\rho} + s^{\rho})(s_{\nu} - r_{\nu}) = \nonumber \eeq
 \beq = \Lambda (s) - \Lambda (r) + A_{0 \mu} (s^{\mu} -r^{\mu}) - \frac{1}{4} F_{\mu \rho} (r^{\rho} + s^{\rho})(s^{\mu} - r^{\mu}) = \nonumber \eeq 
\beq =\Lambda (s) - \Lambda (r) + A_{0 \mu} (s^{\mu} -r^{\mu}) - \frac{1}{4} F_{\mu \nu} (s^{\mu} - r^{\mu})(r^{\nu} + s^{\nu}) \eeq 
Due to antisymmetry of $F_{\mu \nu}$, we know that 
\beq F_{\mu \nu} r^{\mu} r^{\nu} = F_{\mu \nu} s^{\mu} s^{\nu} =0 \eeq
\beq F_{\mu \nu} r^{\mu} s^{\nu} = - F_{\mu \nu} s^{\mu} r^{\nu} \eeq 
and, therefore, 
\beq a (r,s)= \Lambda (s) - \Lambda (r) + A_{0 \mu} (s^{\mu} - r^{\mu}) - \frac{1}{4} F_{\mu \nu} (s^{\mu} -r^{\mu})(r^{\nu} +s^{\nu}) = \nonumber \eeq
\beq = \Lambda (s) - \Lambda (r) + A_{0 \mu} (s^{\mu} - r^{\mu}) + \frac{1}{4} F_{\mu \nu} r^{\mu} s^{\nu} - \frac{1}{4} F_{\mu \nu} s^{\mu} r^{\nu} = \nonumber \eeq
\beq = \Lambda (s) - \Lambda (r) + A_{0 \mu} (s^{\mu} - r^{\mu}) + (1+1) \frac{1}{4} F_{\mu \nu} r^{\mu} s^{\nu} = \nonumber \eeq
\beq = \Lambda (s) - \Lambda (r) + A_{0 \mu} (s^{\mu} - r^{\mu}) + \frac{1}{2} F_{\mu \nu} r^{\mu} s^{\nu} \eeq 
Finally, let us substitute the above holonomies into Eq \ref{FluxIntro} to obtain flux around the loop.
\beq \Phi_{r_1 \cdots r_n r_1} = a(r_n, r_1) + \sum_{k=1}^{n-1} a(r_k, r_{k+1}) = \Lambda (r_1) - \Lambda (r_n) + \sum_{k=1}^{n-1} (\Lambda (r_{k+1}) - \Lambda (r_k)) + \nonumber \eeq 
\beq + A_{0 \mu} \bigg(r_1^{\mu} - r_n^{\mu} + \sum_{k=1}^{n-1} (r_{k+1}^{\mu} -r_k^{\mu}) \bigg) + \frac{1}{2} F_{\mu \nu} \bigg( r_n^{\mu} r_1^{\nu} + \sum_{k=1}^{n-1} r_k^{\mu} r_{k+1}^{\nu} \bigg) = \nonumber \eeq 
\beq = \Lambda (r_1) - \Lambda (r_n) + \Lambda (r_n) - \Lambda (r_1) + A_{0 \mu} (r_1^{\mu} - r_n^{\mu} + r_n^{\mu} - r_1^{\mu}) + \frac{1}{2} F_{\mu \nu} \bigg(r_n^{\mu} r_1^{\nu} + \sum_{k=1}^{n-1} r_k^{\mu} r_{k+1}^{\nu} \bigg) = \nonumber \eeq 
\beq =  \frac{1}{2} F_{\mu \nu} \bigg(r_n^{\mu} r_1^{\nu} + \sum_{k=1}^{n-1} r_k^{\mu} r_{k+1}^{\nu} \bigg) \label{Flux} \eeq
As we have just seen, in computing flux all the terms other than $F_{\mu \nu}$ have cancel out, which makes the flux a function solely of $F_{\mu \nu}$, as expected. 

\subsection*{Electric contribution to electromagnetic Lagrangian}

Let us now proceed to computing Lagrangian density. As we have mentioned in the Introduction, we are looking at the two collections of contours, $C_1$ and $C_2$, where $C_1$ consists of contours of the form $p \prec r \prec q \succ s \succ p$ and $C_2$ consists of contours of the form $p \prec r \prec s \prec q \succ r$ where, in both cases, the only points $x$ satisfying $p \prec x \prec q$ are $r$ and $s$. We have also said that we were looking at sub-collections of such contours for which $q^{\mu} - p^{\mu}$ are parallel to $v^{\mu}$, and which are close to each other coordinate-wise in a reference frame defined by $v^{\mu}$. We would then arrive at the expression of the form $C F_{00}^2 + D F_{ij} F_{ij}$ in a reference frame where $t$-axis coincides with $v^{\mu}$, and our task is to adjust $A/B$ in such a way that we would obtain $D=-C$. As the reader will shortly see, it turns out that the contours that are part of the collection $C_1$ lead to $F_{00}^2$ alone in that frame, while the contours that are part of the collection $C_2$ lead to some linear combination between $F_{00}^2$ and $F_{ij} F_{ij}$. That is why we will call the contribution of contours of $C_1$ ``electric" and the contribution of contours of $C_2$ we will call ``other terms", which is neither electric nor magnetic but combination of both. We will devote this section to said ``electric" contribution (coming from $C_1$), and then we will compute the other terms (coming from $C_2$) in the next section. 

As we mentioned earlier, we are looking at collection of several different elements of $C_1$. Let us consider a sub-collection for which the distance between $p$ and $q$ is $\tau$ (and then we can add up different values of $\tau$ at the end). In this case we can ``superimpose" the contours $p_1 \prec r_1 \prec q_1 \succ s_1 \succ p_1$, $p_2 \prec r_2 \prec q_2 \succ s_2 \succ p_2$ all the way through $p_n \prec r_n \prec q_n \succ s_n \succ p_n$ into imaginary set with a \emph{single} $p$ and \emph{single} $q$ and $2n$ points $\{r_1, \cdots, r_n \}$ and $\{s_1, \cdots, s_n \}$, all of which are satisfying $p \prec r_k \prec q$ and $p \prec s_k \prec q$. The reason we can't superimpose $r_k$ and $s_k$ is that, \emph{if we assume the manifold background}, their \emph{continuum} distance to $p$ and $q$ would alter. Now, the way it reflects on the discrete setting is that, if we assume that $A^{\mu}$ is linear in continuum coordinates, then the values of $a (p,r)$, $a(r,q)$, $a (q,r)$, $a(r,p)$ and $a(r,s)$ will all depend on the location of $r$ and $s$ \emph{in a continuum}. Therefore, even though the contours $p_1 \prec r_1 \prec q_1 \succ s_1 \succ p_1$ and $p_2 \prec r_2 \prec q_2 \succ s_2 \succ p_2$ look identical combinatorially, they can't be superimposed since their contributions to flux are different. However $p$-s and $q$-s \emph{can} be superimposed, since we made sure that \emph{continuum} distance between $p$ and $q$ is the same in both cases (and we will integrate over different continuum distances between $p$ and $q$ at the end). 

Anyway, after we superimposed contours as described, the contour $pr_kqs_kp$ is no longer an element of $C_1$: after all, part of the definition of $C_1$ is that $r$ and $s$ are the \emph{only} elements satisfying $p \prec x \prec q$; but, in this case, we have $2n$ different elements satisfying this condition. Instead, we will introduce a structure, called \emph{Alexandrov set} 
\beq \alpha (p,q) = \{ x \vert p \prec x \prec q \} \eeq
and simply say $r_k \in \alpha (p,q)$ and $s_k \in \alpha (p,q)$. Strictly speaking what we said so far is that $\alpha (p,q)$ has $2n$ elements, namely $\{r_1, \cdots, r_n \} \cup \{s_1, \cdots, s_n \}$ and in each contour we select one element out of $\{r_1, \cdots, r_n \}$ and the other element out of $\{s_1, \cdots, s_n \}$. Such partition into two subsets, however, is not necessary. After all, if $n$ is large enough, then each subset is expected to distribute uniformly throughout the interior of $\alpha (p,q)$, which means that if we select $r$ and $s$ from the same subset rather than two different ones we would statistically expect the same answer and, therefore, if we simply relax the condition as to which subset $r$ and $s$ are selected from, our answer will simply multiply by overall factor $4$, which we aren't worried about. This being the case, we can relax the condition that the number of elements of $\alpha (p,q)$ is even, and simply write
\beq {\cal L}_{Electric} (a; p,q) = \sum_{r, s \in \alpha (p,q)} \Phi^2 (prsqp) \eeq 
and that sum will then be replaced with an integral,
\beq {\cal L}_{Electric} (a; p,q) = \int_{\alpha (p,q)} d^d r d^d s \; \Phi^2 (prsqp) \eeq 
 In order to evaluate this integral,  let us assume that the Lorentzian distance between $p$ and $q$ is $\tau$, and let us set a coordinate system in such a way that $t$-axis passes through $p$ and $q$, with origin in the middle. Thus, in Cartesian coordinates,  
\beq p = \bigg(- \frac{\tau}{2}, 0, \cdots, 0 \bigg) \; , \; q = \bigg(\frac{\tau}{2}, 0 \cdots, 0 \bigg) \eeq 
Therefore, we have 
\beq a(p, r) = \frac{1}{2} F_{\mu \nu} p^{\mu} r^{\nu} + A_{0 \mu} (r^{\mu} - p^{\mu}) + \Lambda (r) - \Lambda (p)= \nonumber \eeq
 \beq =^{p^k=0} \frac{1}{2} F_{0 \nu} p^0 r^{\nu} + A_{0 \mu} (r^{\mu} - p^{\mu}) + \Lambda (r) - \Lambda (p) = \nonumber \eeq 
\beq =^{F_{00}=0} \frac{1}{2} F_{0k} p^0 r^k + A_{0 \mu} (r^{\mu} - p^{\mu}) + \Lambda (r) - \Lambda (p) = \nonumber \eeq
 \beq = \frac{1}{2} F_{0k} \bigg(- \frac{\tau}{2} \bigg) r^k + A_{0 \mu} (r^{\mu} - p^{\mu}) + \Lambda (r) - \Lambda (p) = \nonumber \eeq
 \beq = - \frac{\tau}{4} F_{0k} r^k + A_{0 \mu} (r^{\mu} - p^{\mu}) + \Lambda (r) - \Lambda (p) \eeq
Similarly, 
\beq a(r, q) = \frac{1}{2} F_{\mu \nu} r^{\mu} q^{\nu} + A_{0 \mu} (q^{\mu} - r^{\mu}) + \Lambda (q) - \Lambda (r) = \nonumber \eeq
 \beq =^{q^k=0}  \frac{1}{2}F_{\mu 0}r^{\mu} q^0 + A_{0 \mu} (q^{\mu} - r^{\mu}) + \Lambda (q) - \Lambda (r) = \nonumber \eeq
 \beq =^{F_{00}=0} \frac{1}{2} F_{k0} r^k q^0 + A_{0 \mu} (q^{\mu} - r^{\mu}) + \Lambda (q) - \Lambda (r) = \nonumber \eeq
 \beq = \frac{1}{2} F_{k0} r^k \bigg(\frac{\tau}{2} \bigg) + A_{0 \mu} (q^{\mu} - r^{\mu}) + \Lambda (q) - \Lambda (r) = \nonumber \eeq 
\beq = \frac{\tau}{4} F_{k0} r^k + A_{0 \mu} (q^{\mu} - r^{\mu}) + \Lambda (q) - \Lambda (r)  \eeq 
and, finally, 
\beq a(p,q) = \frac{1}{2} F_{\mu \nu} p^{\mu} q^{\nu} + A_{0 \mu} (q^{\mu} - p^{\mu}) + \Lambda (q) - \Lambda (p) = \nonumber \eeq
\beq =^{p^k=q^k=0} \frac{1}{2} F_{00}p^0q^0 + A_{0 \mu} (q^{\mu} - p^{\mu}) + \Lambda (q) - \Lambda (p) = \nonumber \eeq
\beq =^{F_{00}=0}  A_{0 \mu} (q^{\mu} - p^{\mu}) + \Lambda (q) - \Lambda (p) = \eeq 
Thus, the flux through a contour $prqp$ is 
\beq \Phi_{prqp} =  - \frac{\tau}{4} F_{0k} r^k + A_{0 \mu} (r^{\mu} - p^{\mu}) + \Lambda (r) - \Lambda (p) + \nonumber \eeq
\beq + \frac{\tau}{4} F_{k0} r^k + A_{0 \mu} (q^{\mu} - r^{\mu}) + \Lambda (q) - \Lambda (r)  + \nonumber \eeq
\beq + A_{0 \mu} (p^{\mu} - q^{\mu}) + \Lambda (p) - \Lambda (q) = \nonumber \eeq 
\beq =- (1+1) \frac{\tau}{4} F_{0k} r^k + A_{0 \mu} (r^{\mu} -p^{\mu}+q^{\mu}-r^{\mu} + p^{\mu} - q^{\mu}) + \nonumber \eeq
 \beq + \Lambda (r) - \Lambda (p) + \Lambda (q) - \Lambda (r) + \Lambda (p) - \Lambda (q) = \nonumber \eeq 
\beq = - 2 \frac{\tau}{4} F_{0k} r^k = - \frac{\tau}{2} F_{0k} r^k \eeq
and the flux through four-contour is 
\beq \Phi_{prqsp} = \Phi_{prqp} + \Phi_{pqsp} = \Phi_{prqp} - \Phi_{psqp} = - \frac{\tau}{2} F_{0k} r^k - \bigg(- \frac{\tau}{2} F_{0k} s^k \bigg) = \frac{\tau}{2} F_{0k} (s^k - r^k) \eeq
Obviously, this will only give us an electric contribution, so we would need other terms to get magnetic contribution, which we will talk about from next section onward. But for now lets finish the calculation of electric contribution. For the dimension $d$ and an integer $k$ lets define $I_{dk}$ to be the scalar coefficient in the equation
\beq \int_{\alpha (p,q)} d^d r \; (r^1)^k = I_{dk} \tau^{d+k} \eeq
where $r^1$ can be replaced with $r^2$ or any other spacelike coordinate, and $\tau^{d+k}$ follows from dimensional analysis. We will explicitly compute $I_{dk}$ in the future sections, but for now let us go ahead with the calculation, assuming that we know what its equal to. From symmetry considerations, we know that 
\beq k \; {\rm is \; odd \;} \Longrightarrow \; I_{dk}=0 \eeq
It turns out that the integration over the flux through square loop $p \prec r \prec q \succ s \succ p$ is easier if we break it into the two triangles, $p \prec r \prec q \succ p$ and $p \prec q \succ s \succ p$. The flux around the \emph{triangular} contour is
\beq \int_{\alpha (p,q)} d^d r \; (a (p,r) + a (r,q) + a (q,p))^2 = \int_{\alpha (p,q)} d^d r \; \bigg(- \frac{\tau}{2} F_{0k} r^k \bigg) \bigg(- \frac{\tau}{2} F_{0l} r^l \bigg) = \nonumber \eeq
\beq = \frac{\tau^2}{4} F_{0k} F_{0l} \int_{\alpha (p,q)} d^d r \; r^k r^l = \bigg(\frac{\tau^2}{4} F_{0k} F_{0l} \bigg) (\delta^k_l I_{d2} \tau^{d+2}) = \frac{1}{4} I_{d2}  \tau^{d+4} F_{0k} F_{0k} \label{ElectricTriangle} \eeq
and, if we instead pick square loop, we obtain
\beq \int_{\alpha (p,q)} d^d r \; (a (p,r) + a (r,q) + a (q,s) + a (s,p))^2 = \nonumber \eeq
 \beq = \int_{r,s \in \alpha (p,q)} d^d r d^d s \; \bigg(\frac{\tau}{2} F_{0k} (s^k-r^k) \bigg) \bigg( \frac{\tau}{2} F_{0l} (s^l-r^l) \bigg) = \nonumber \eeq
\beq = \frac{\tau^2}{4} F_{0k} F_{0l} \int_{r,s \in \alpha (p,q)} d^d r d^d s \; (s^ks^l - s^kr^l - r^ks^l +r^kr^l) = \nonumber \eeq 
\beq = \frac{\tau^2}{4} F_{0k} F_{0l} \bigg(\bigg(\int_{\alpha (p,q)} d^d r \bigg) \bigg(\int_{\alpha (p,q)} d^d s \; s^k s^l \bigg) - \bigg(\int_{\alpha (p,q)} d^d s \; s^k \bigg) \bigg( \int_{\alpha (p,q)} d^d r \; r^l \bigg) - \nonumber \eeq 
\beq - \bigg(\int_{\alpha (p,q)} d^d r \; r^k \bigg) \bigg(\int_{\alpha (p,q)} d^d s \; s^l \bigg) + \bigg( \int_{\alpha (p,q)} d^d r \; r^k r^l \bigg) \bigg( \int_{\alpha (p,q)} d^d s \bigg) \bigg)  \eeq 
Now if we recall that we picked the origin to be in the center of Alexandrov set, we know that the integrals of odd functions give zero. Thus, the above expressing evaluates to 
\beq \int_{\alpha (p,q)} d^d r \; (a (p,r) + a (r,q) + a (q,s) + a (s,p))^2 = \nonumber \eeq
\beq =  \frac{\tau^2}{4} F_{0k} F_{0l} \Big( \big(I_{d0} \tau^d \big) \big(I_{d2} \tau^{d+2} \delta^k_l \big) - 0*0 -0*0 + \big(I_{d2} \tau^{d+2} \delta^k_l \big) \big(I_{d0} \tau^d) \Big) = \nonumber \eeq 
\beq = \bigg(\frac{\tau^2}{4} F_{0k} F_{0l} \bigg) \big(2I_{d0}I_{d2} \tau^{2d+2} \delta^k_l \big) = \frac{\tau^2 \tau^{2d+2}}{4} 2 I_{d0}I_{d2} \delta^k_l F_{0k} F_{0l} = \frac{I_{d0} I_{d2}}{2}  \tau^{2d+4} F_{0k} F_{0k} \label{ElectricSquare} \eeq
where we were dropping the integrals of odd terms. In both cases we obtain $F_{0k} F_{0k}$, its simply that the coefficients are different. This would be electric part of the Lagrangian in the reference frame defined by endpoints of Alexandrov set, so we need something else for the magnetic part. 

\subsection*{Other terms in the Lagrangian} 
 
As we talked about earlier, we are considering two types of contours. The contours of the form $p \prec r \prec q \succ s \succ p$ produce purely electric contribution, while the contours of the form $p \prec r \prec s \prec q \succ p$ produce a linear combination of electric and magnetic contributions, and the linear combination of both with appropriately adjusted ratio of coefficients, $A/B$, would produce $F^{\mu \nu} F_{\mu \nu}$ we are seeking. In the previous sectoin we were exclusively focusing on $p \prec r \prec q \succ s \succ p$ contours. In the upcoming section, let us look at $p \prec r \prec s \prec q \succ p$ ones. By the same argument as before, we will superimpose $p$-s and $q$-s of different contours into a single $p$ and $q$, while integrating over different choices of $r$ and $s$, which is why, once again, $\alpha (p,q)$ will contain multiple points. Without further due, let us proceed with the calculation. We know from Eq \ref{Flux} that 
\beq \Phi (r_1 \cdots r_n r_1)= \frac{1}{2} F_{\mu \nu} \bigg(r_n^{\mu} r_1^{\nu} + \sum_{k=1}^{n-1} r_k^{\mu} r_{k+1}^{\nu} \bigg) \eeq
If we now substitute $(prsqp)$ in place of $(r_1 \cdots r_n r_1)$, we would have $4$ in place of $n$ and, therefore, we would have the sum of four terms, it is also easy to see that each term is either zero-th or first order in each $r$ and $s$. Now, if we look at $\Phi^2$ rather than $\Phi$, then we will have all possible products of pairs of those terms. Since each term contains $r$ either zero or one time, the product of two such terms can contain $r$ either $0+0=0$, or $0+1=1$ or $1+0=1$ or $1+1=2$ times. In particular, the expression we obtain is the following: 
\beq \int \Phi_{prsqp}^2 d^d r d^d s = \nonumber \eeq
\beq = \int d^d r d^d s \; \bigg(\frac{1}{4} F_{\alpha \beta} F_{\gamma \delta} r^{\alpha} s^{\beta} r^{\gamma} s^{\delta} + \frac{\tau^2}{16} F_{0i} F_{0j} r^i r^j + \frac{\tau^2}{16} F_{0i} F_{0j} s^i s^j + \nonumber \eeq
\beq + \frac{\tau^2}{4} F_{0i} F_{0j} r^i s^j + \frac{\tau^2}{4} F_{0i} F_{0j} s^i r^j \bigg)  \eeq 
If we plug in $p^0= - \tau/2$, $q^0 = \tau/2$ and $p^k=q^k=0$, this becomes 
\beq \int \Phi_{prsqp}^2 d^d r d^d s = \nonumber \eeq
\beq = \int d^d r d^d s \; \bigg(\frac{1}{4} F_{\alpha \beta} F_{\alpha \beta} r^{\alpha} s^{\beta} r^{\alpha} s^{\beta} + \frac{1}{4} F_{\alpha \beta} F_{\beta \alpha} r^{\alpha} s^{\beta} r^{\beta} s^{\alpha} + \frac{\tau^2}{16} F_{0k} F_{0k} (r^k)^2  + \nonumber \eeq
\beq + \frac{\tau^2}{16} F_{0k} F_{0k} (s^k)^2+ \frac{\tau^2}{4} F_{0i} F_{0j} r^is^j + \frac{\tau^2}{4} F_{0i} F_{0j} s^is^j \bigg) \eeq 
Now, in place of each of each term with one Greek index, we will put two terms: namely where that Greek index is zero (time component) and where that Greek index is replaced with Latin index (space component). In case of product of two Greek indexes, we will obtain $2 \times 2 = 4$ terms: namely where they are both zero, where they are both Latin indexes, where first one is zero and second one is Latin, and where first one is Latin and second one is zero. Thus, we obtain
\beq \int \Phi_{prsqp}^2 d^d r d^d s = \nonumber \eeq
\beq = \int d^d r d^d s \; \bigg(\frac{1}{2} F_{0k} F_{0k} r^0s^kr^0s^k + \frac{1}{4} F_{k0} F_{k0} r^k s^0 r^k s^0 + \nonumber \eeq 
\beq + \frac{1}{4} F_{ij} F_{ij} r^is^jr^is^j + \frac{1}{4} F_{0k} F_{k0} r^0 s^k r^k s^0 + \nonumber \eeq
\beq + \frac{1}{4} F_{k0} F_{0k} r^k s^0 r^0 s^k + \frac{1}{4} F_{ij} F_{ji} r^iS^jr^js^i + \nonumber \eeq
\beq + \frac{\tau^2}{16} F_{0k} F_{0k} (r^k)^2 + \frac{\tau^2}{16} F_{0k} F_{0k} (s^k)^2 + \nonumber \eeq
\beq + \frac{\tau^2}{4} F_{0i} F_{0j} r^is^j + \frac{\tau^2}{16} F_{0i} F_{0j} r^j s^i  \bigg) = \nonumber \eeq 
\beq = \int d^d r d^d s \; \bigg(\frac{1}{4} F_{0k} F_{0k} (r^0)^2 (s^k)^2 + \frac{1}{4} F_{0k} F_{0k} (r^k)^2 (s^0)^2 + \nonumber \eeq
\beq + \frac{1}{4}F_{ij} F_{ij} (r^i)^2(s^j)^2 - \frac{1}{4} F_{0k} F_{0k} r^0s^0 r^ks^k - \nonumber \eeq 
\beq - \frac{1}{4} F_{0k} F_{0k} r^0 s^0 r^k s^k - \frac{1}{4} F_{ij} F_{ij} r^ir^j s^is^j + \nonumber \eeq 
\beq + \frac{\tau^2}{16} F_{0k} F_{0k} (r^k)^2 + \frac{\tau^2}{16} F_{0k} F_{0k} (s^k)^2 + \frac{\tau^2}{4} F_{0i} F_{0j} (r^is^j+r^js^i) \eeq 
If we now pull constants out of the integral, and combine integrals whose constant coefficients are the same, we obtain
\beq \int \Phi_{prsqp}^2 d^d r d^d s = \nonumber \eeq
\beq = \frac{1}{4} F_{0k} F_{0k} \int d^d r d^d s \; \bigg((r^0)^2 (s^k)^2 + (r^k)^2 (s^0)^2 - r^0s^0 r^k s^k - \nonumber \eeq 
\beq - r^0 s^0 r^k s^k + \frac{\tau^2}{4} (r^k)^2 + \frac{\tau^2}{4} (s^k)^2 + 2 \tau^2 r^k s^k \bigg) + \nonumber \eeq 
\beq + \frac{\tau^2}{4} F_{ij} F_{ij} \int d^d r d^d s \; ((r^i)^2 (s^j)^2 - r^i r^j s^i s^j)  \eeq 
By inspecting the above expression one can see that some of the therms are the same. By combining same terms, we obtain
\beq \int \Phi_{prsqp}^2 d^d r d^d s = \nonumber \eeq
\beq = \frac{1}{4} F_{0k} F_{0k} \int d^d r d^d s \bigg(2(r^0)^2 (s^k)^2 - 2r^0s^0 r^k s^k + 2 \bigg(\frac{\tau^2}{4}\bigg) (r^k)^2 + 2 \tau^2 r^k s^k \bigg) + \nonumber \eeq
\beq + \frac{1}{4} F_{ij} F_{ij} \int d^d r d^d s ((r^i)^2(s^j)^2 - r^ir^js^is^j)  \eeq
\beq = \frac{1}{2} F_{0k} F_{0k} \int d^d r d^d s \; \bigg((r^0)^2 (s^k)^2 - r^0 s^0 r^k s^k + \frac{\tau^2}{4} (r^k)^2 + \tau^2 r^k s^k \bigg) + \nonumber \eeq
\beq + \frac{1}{4} F_{ij} F_{ij} \int d^d r d^d s ((r^i)^2 (s^j)^2 - r^ir^j s^is^j) \label{SumOfDoubleIntegrals}\eeq 
Notably, \emph{despite} origin being in the center of Alexandrov set, we did \emph{not} drop out some of the terms that \emph{look like} they are odd. The reason for this is that $r$ and $s$ are \emph{not} independent of each other, thanks to the relation $r \prec s$. Now, since $s$ is being integrated over $\alpha (r,q)$ as opposed to $\alpha (p,q)$, the center of $\alpha (r,q)$ is no longer at the origin and, therefore, it is no longer true that the integral of odd function of $s$ over $\alpha (r,q)$ is zero. On the contrary, such integral will be a function of $r$ since the location of the center of $\alpha (r,q)$ will be. As a result, we might get an expression of the form 
\beq \int_{p \prec r \prec s \prec q} d^d r d^d s \; r^1 s^1 = \int_{\alpha (p,q)} \bigg( d^d r \; r^1 \int_{\alpha(r,q)} d^d s \;  s^1 \bigg) = \nonumber \eeq
\beq =  \int_{\alpha (p,q)} d^d r \; r^1 (kr^1 + \cdots) =  \int_{\alpha (p,q)} (k(r^1)^2 + \cdots) \neq 0 \label{FromDoubleToSingleOutline} \eeq
By similar argument, the integrals of $r^0s^0$ and some other ones return non-zero answers as well. 

\subsection*{Re-expression double integrals in terms of single integrals}

From this point on, the main thing we have to accomplish is to evaluate the integrals in Eq \ref{SumOfDoubleIntegrals}. As we pointed out, part of what makes it complicated is that those are double integrals. Therefore, in this section we will reduce them to single integrals and then in the sections that follow we will try and evaluate corresponding single integrals. For the purposes of this section, we will simply assume that the single integrals take the form
\beq \int_{\alpha (p,q)} d^d r = I_{d0} \tau^d (p,q) \label{Setup1} \eeq
\beq \int_{\alpha (p,q)} d^d r \; (r^1)^2 = I_{d2} \tau^{d+2} (p,q) \label{Setup2} \eeq 
\beq \int_{\alpha (p,q)} d^d r \; (r^0)^2 = J_{d2} \tau^{d+2} (p,q) \label{Setup3} \eeq 
where calculations leading to the explicit expressions for the coefficients $I_{d0}$, $I_{d2}$ and $J_{d2}$ are postponed to future sections. The way we obtain single integrals from the double integral is exemplified in Eq \ref{FromDoubleToSingleOutline}. Without further due, let us proceed with six calculations each intended to evaluate six corresponding terms in Eq \ref{SumOfDoubleIntegrals}. 

{\bf Calculation 1} 

Following the outline of Eq \ref{FromDoubleToSingleOutline}, we write 
\beq \int_{p \prec r \prec s \prec q} d^d r d^d s (r^1 s^1)= \int_{p \prec r \prec q} \bigg(d^d r \; r^1 \int_{r \prec s \prec q} d^d s \; s^1 \bigg)  \eeq
For reasons that will be clear shortly, let us rewrite it as 
\beq \int_{p \prec r \prec s \prec q} d^d r d^d s (r^1 s^1) = \int_{p \prec r \prec q} \bigg(d^d r \; r^1 \int_{r \prec s \prec q} d^d s \;  \bigg(\bigg(s^1 - \frac{r^1+q^1}{2}\bigg)+\frac{r^1+q^1}{2} \bigg) \bigg)  \eeq 
Now we note that $s^1 - (r^1+q^1)/2$ is symmetric about the center of Alexandrov set and, therefore, integrates to zero. One might want to be a bit careful since the axis of $\alpha (r,q)$ is tilted compared to axis of $\alpha (p,q)$. Thus, as far as the integration around $\alpha (r, q)$ is concerned, one would need new coordinates $s^{\prime \mu}$ rather than $s^{\mu}$. However, if we remember to include linear displacement in the new coordinate system, so that the origin of primed coordinate system coincides with the center of $\alpha (r,q)$ as opposed to $\alpha (p,q)$, we will see that $s^1 - (r^1+q^1)/2$ is a linear combination of $s^{\prime 0}$ and $s^{\prime 1}$, both of which integrate to zero. Thus, we can throw away $s^1 - (r^1+q^1)/2$ and obtain 
\beq  \int_{p \prec r \prec s \prec q} d^d r d^d s (r^1 s^1)= \int_{p \prec r \prec q} \bigg(d^d r \frac{r^1 (r^1+q^1)}{2} \int_{r \prec s \prec q} d^d s \bigg)  \eeq 
Now, to compute integral over $s$, we use  Eq \ref{Setup1}, where, instead of $\tau (p,q)$ we will put $\tau (r,q)$ which, in turn, we express in turn we will express in terms of difference of squares of coordinates. This will produce the following expression: 
\beq \int_{p \prec r \prec s \prec q} d^d r d^d s (r^1 s^1)= \int_{p \prec r \prec q} \bigg(d^d r \; \frac{r^1 (r^1+q^1)}{2} I_{d0} \bigg(\bigg(\frac{\tau}{2} -r^0 \bigg)^2 - \vert \vec{r} \vert^2 \bigg)^{d/2} \bigg)  \eeq 
Now, at this point, since we are integrating over a single variable $r$, the terms that appear to be odd are, in fact, odd, and integrate to zero. By throwing them away, we obtain
\beq \int_{p \prec r \prec s \prec q} d^d r d^d s (r^1 s^1)= \frac{I_{d0}}{2} \int_{p \prec r \prec q} d^d r \; \bigg[ (r^1)^2 \bigg( \bigg(\frac{\tau}{2} -r^0 \bigg)^2 - \vert \vec{r} \vert^2 \bigg)^{d/2} \bigg] \eeq 

{\bf Calculation 2} 

Using similar techniques as previously, 
\beq \int_{p \prec r \prec s \prec q} d^d r d^d s \; (r^1)^2 = \int_{p \prec r \prec q} \bigg(d^d r \; (r^1)^2 \int_{r \prec s \prec q} d^d s \bigg) = \nonumber \eeq 
\beq = \int_{p \prec r \prec q} \bigg(d^d r \; (r^1)^2 I_{d0} \bigg(\bigg(\frac{\tau}{2}-r^0\bigg)^2-\vert \vec{r} \vert^2 \bigg)^{d/2} \bigg) = \nonumber \eeq 
\beq = I_{d0}  \int_{p \prec r \prec q} \bigg(d^d r \; (r^1)^2  \bigg(\bigg(\frac{\tau}{2}-r^0\bigg)^2-\vert \vec{r} \vert^2 \bigg)^{d/2} \bigg) = \nonumber \eeq
\beq = 2 \times ({\rm previous \; answer}) \eeq 

{\bf Calculation 3}

\beq \int_{p \prec r \prec s \prec q} d^d r d^d s \; (r^1)^2 (s^2)^2 = \int_{p \prec r \prec q} \bigg(d^d r \; (r^1)^2 \int_{r \prec s \prec q} d^d s \; (s^2)^2 \bigg)  \eeq 
If we were to have $(s^1)^2$, then we would have to express $s^1$ in terms of a linear combination of $s^{\prime 0}$ and $s^{\prime 1}$ and then evaluate two separate integrals. However, since we have $s^2$ rather than $s^1$, and the boost from $pq$-frame to $rq$-frame occurs in $01$-plane, we know that $s^{\prime 2} = s^2$ and therefore, we can use Eq \ref{Setup2}. But, as before, we have to use the distance from $r$ to $q$ rather than from $p$ to $q$ in our formula, which we again write in terms of difference of squares of coordinates. Thus we obtain 
\beq \int_{p \prec r \prec s \prec q} d^d r d^d s \; (r^1)^2 (s^2)^2 = \int_{p \prec r \prec q} \bigg(d^d r \; (r^1)^2 I_{d2} \bigg(\bigg(\frac{\tau}{2}-r^0\bigg)^2 - \vert \vec{r} \vert^2 \bigg)^{1+ \frac{d}{2} }\bigg) = \nonumber \eeq 
\beq = I_{d2} \int_{p \prec r \prec q} \bigg(d^d r \; (r^1)^2 \bigg(\bigg(\frac{\tau}{2}-r^0\bigg)^2 - \vert \vec{r} \vert^2 \bigg)^{1+ \frac{d}{2} }\bigg) \eeq 

{\bf Calculation 4}

\beq \int_{p \prec r \prec s \prec q} d^d r d^d s (r^0)^2 (s^1)^2 = \int_{p \prec r \prec q} \bigg(d^d r \; (r^0)^2 \int_{p \prec r \prec q} d^d s (s^1)^2 \bigg) \label{Calc4Start} \eeq
As we have just pointed out, now that we have $s^1$ rather than $s^2$ we do, in fact, have to carry out Lorentz transformation. So we would like to go to the reference frame of an observer that moves with constant velocity from point $r$ to point $q$. The velocity of that observer, with respect to original oberver, is 
\beq v^{\mu} = \frac{q^{\mu}-r^{\mu}}{\big(\big(\frac{\tau}{2}-r^0\big)^2- \vert \vec{r} \vert^2 \big)^{1/2}} \eeq 
In light of the fact that $v^{\mu} v_{\mu}=1$, we know that in the reference frame of the new observer $v^0$ coincides with $1$. Therefore, in the reference frame of the old observer, $v^0$ is equal to $1$ multiplied by Lorentz factor $\gamma$:
\beq v^0 = 1 \gamma = \gamma \eeq
and therefore, we can immediately read off what $\gamma$ is: 
\beq  \gamma = v^0 = \frac{\frac{\tau}{2}-r^0}{\big(\big(\frac{\tau}{2}-r^0\big)^2- \vert \vec{r} \vert^2 \big)^{1/2}} \eeq
Now, in order to write down $s^1$ in terms of $s^{\prime 1}$ and $s^{\prime 0}$, we have to keep in mind that, apart from Lorentz boost, there is also a translation: in particular, in original coordinate system the origin is between $p$ and $q$, while in new coordinate system the origin is between $r$ and $q$. So, keeping in mind both Lorentz boost and the translation, we have
\beq s^1 = \gamma \bigg(s^{\prime 1} + \frac{v^1}{v^0} s^{\prime 0} \bigg) + \frac{q^1+r^1}{2} = \frac{\frac{\tau}{2}-r^0}{\big(\big(\frac{\tau}{2}-r^0\big)^2- \vert \vec{r} \vert^2 \big)^{1/2}} \bigg(s^{\prime 1} - \frac{r^1 s^{\prime 0}}{\frac{\tau}{2}-r^0} \bigg) + \frac{r^1}{2} \label{SpaceTransformOfS} \eeq
Since the origin of the primed coordinate system is at the center of $\alpha (r,q)$, all of the odd terms in the primed coordinate system integrate to zero. Therefore, when we take the square of $s^1$, we will simply add squares of each term, without any cross terms: 
\beq \int_{p \prec r \prec s \prec q} d^d r d^d s (r^0)^2 (s^1)^2 = \int_{p \prec r \prec q} \bigg(d^d r \; (r^0)^2 \int_{p \prec r \prec q} d^d s (s^1)^2 \bigg) = \nonumber \eeq
\beq = \int_{p \prec r \prec q} \bigg(d^d r \; (r^0)^2 \int_{r \prec s \prec q} d^d s' \bigg(\frac{\big(\frac{\tau}{2} -r^0\big)^2}{\big(\frac{\tau}{2}-r^0\big)^2- \vert \vec{r} \vert^2} \bigg((s^{\prime 1})^2 + \frac{(r^1)^2}{\big(\frac{\tau}{2}-r^0\big)^2} (s^{\prime 0})^2 \bigg) + \frac{(r^1)^2}{4} \bigg) \bigg)  \eeq
Since the above result is written in primed coordinate, we don't need to worry about transformations any more and we can go ahead and substitute the expressions for the integrals (while of course keeping in mind the appropriate distance): 
\beq \int_{p \prec r \prec s \prec q} d^d r d^d s (r^0)^2 (s^1)^2 =  \nonumber \eeq 
\beq = \int_{p \prec r \prec q} \bigg\{ d^d r \; (r^0)^2 \bigg\{\frac{\big(\frac{\tau}{2}-r^0\big)^2}{\big(\frac{\tau}{2} -r^0 \big)^2 - \vert \vec{r} \vert^2} \bigg[I_{d2} \bigg(\bigg(\frac{\tau}{2}-r^0\bigg)^2- \vert \vec{r} \vert^2 \bigg)^{1+\frac{d}{2}}- \nonumber \eeq
 \beq +  \frac{(r^1)^2}{\big(\frac{\tau}{2}-r^0\big)^2} J_{d2} \bigg(\bigg(\frac{\tau}{2} - x^0 \bigg)^2 - \vert \vec{r} \vert^2 \bigg)^{1+\frac{d}{2}} \bigg] + \frac{(r^1)^2}{4} I_{d0} \bigg[\bigg(\frac{\tau}{2}-r^0\bigg)^2 - \vert \vec{r} \vert^2 \bigg]^{d/2} \bigg\} \bigg\}  \eeq 
Now, it happens that all of the terms with the power of $1+ (d/2)$ have the power of $1$ in the denominator. Thus, after cancellation, all terms have power of $d/2$: 
\beq \int_{p \prec r \prec s \prec q} d^d r d^d s (r^0)^2 (s^1)^2 =  \nonumber \eeq 
\beq = \int_{p \prec r \prec q} \bigg(d^d r (r^0)^2 \bigg(I_{d2} \bigg(\frac{\tau}{2}-r^0\bigg)^2 \bigg(\bigg(\frac{\tau}{2}-r^0 \bigg)^2 - \vert \vec{r} \vert^2 \bigg)^{d/2} + J_{d2} (r^1)^2 \bigg(\bigg(\frac{\tau}{2}-r^0\bigg)^2 - \vert \vec{r} \vert^2 \bigg)^{d/2} + \nonumber \eeq
\beq + \frac{I_{d0}}{4} (r^1)^2 \bigg(\bigg(\frac{\tau}{2}-r^0\bigg)^2 -\vert \vec{r} \vert^2 \bigg)^{d/2} \bigg) \bigg)  \eeq
And, finally, we split that integral into pieces, to obtain  
\beq \int_{p \prec r \prec s \prec q} d^d r d^d s (r^0)^2 (s^1)^2 =  \nonumber \eeq 
\beq = I_{d2} \int_{p \prec r \prec q} d^d r \; (r^0)^2 \bigg(\frac{\tau}{2}-r^0\bigg)^2 \bigg(\bigg(\frac{\tau}{2}-r^0\bigg)^2- \vert \vec{r} \vert^2 \bigg)^{d/2} + \nonumber \eeq
\beq + \bigg(\frac{I_{d0}}{4}+J_{d2}\bigg) \int_{p \prec r \prec q} d^d r \; (r^1)^2 (r^0)^2 \bigg(\bigg(\frac{\tau}{2} -r^0 \bigg)^2 - \vert \vec{r} \vert^2 \bigg)^{d/2} \eeq 

{\bf Calculation 5}

\beq \int_{p \prec r \prec s \prec q} d^d r d^d s\; (r^1 s^2)(r^2 s^1) = \int_{p \prec r \prec q} \bigg(d^d r \; r^1 r^2 \int_{r \prec s \prec q} d^d s \; s^1 s^2 \bigg)  \eeq 
Recall that 
\beq s^k = \frac{\frac{\tau}{2}-r^0}{\sqrt{\big(\frac{\tau}{2} - r^0 \big)^2 - \vert \vec{r} \vert^2}} \bigg(s^{\prime k} - \frac{r^k s^{\prime 0}}{\frac{\tau}{2} - r^0} \bigg)+ \frac{r^k}{2} \label{skLawrentz}\eeq 
In the previous case, when we were looking at $s^1 s^1$, we had an $s^{\prime 1} s^{\prime 1}$ term, that had non-zero contribution. This time, however, we have $s^{\prime 1} s^{\prime 2}$ term instead, which integrates to zero. And, of course, $s^{\prime k} s^{\prime 0}$ integrates to zero as well, in both cases. Thus, the only non-zero contribution to the integral come from $s^{\prime 0} s^{\prime 0}$ as well as so-called constant $r^k r^k$. Thus, we obtain
\beq \int_{p \prec r \prec s \prec q} d^d r d^d s\; (r^1 s^2)(r^2 s^1) = \int_{p \prec r \prec q} \bigg(d^d r \; r^1 r^2 \int_{r \prec s \prec q} d^d s \; s^1 s^2 \bigg)  = \nonumber \eeq 
\beq = \int_{p \prec r \prec q} \Bigg(d^d r \; r^1 r^2 \int d^d s' \; \Bigg(\frac{\frac{\tau}{2} -r^0}{\sqrt{\big(\frac{\tau}{2}-r^0\big)^2- \vert \vec{r} \vert^2}} \Bigg)^2 \frac{r^1r^2(s^{\prime 0})^2}{\big(\frac{\tau}{2}-r^0\big)^2} + \frac{r^1r^2}{4} \Bigg)  \eeq
After canceling $((\tau/2)-r^0)^2$, we obtain 
\beq \int_{p \prec r \prec s \prec q} d^d r d^d s\; (r^1 s^2)(r^2 s^1) = \nonumber \eeq
\beq = \int_{p \prec r \prec q} \bigg(d^d r \; r^1 r^2 \int d^d s' \; \frac{r^1r^2 (s^{\prime 0})^2}{\big(\frac{\tau}{2}-r^0\big)^2- \vert \vec{r} \vert^2} + \frac{r^1r^2}{4} \bigg)  \eeq 
By evaluating the integrals over $s'$ we obtain
\beq \int_{p \prec r \prec s \prec q} d^d r d^d s\; (r^1 s^2)(r^2 s^1) = \nonumber \eeq 
\beq = \int_{p \prec r \prec q} \bigg( d^d r \; r^1 r^2 \bigg(\frac{r^1r^2}{\big(\frac{\tau}{2}-r^0 \big)^2 - \vert \vec{r} \vert^2} J_{d2} \bigg(\bigg(\frac{\tau}{2}-r^0\bigg)^2- \vert \vec{r}\vert^2\bigg)^{1+\frac{d}{2}} + \nonumber \eeq
\beq + \frac{r^1r^2}{4} I_{d0} \bigg(\bigg(\frac{\tau}{2}-r^0 \bigg)^2- \vert \vec{r} \vert^2 \bigg)^{d/2} \bigg) \bigg) = \nonumber \eeq
and finally we split integral into pieces and obtain
 \beq \int_{p \prec r \prec s \prec q} d^d r d^d s\; (r^1 s^2)(r^2 s^1) = \eeq
\beq = J_{d2} \int_{p \prec r \prec q} d^d r (r^1)^2 (r^2)^2 \bigg(\bigg(\frac{\tau}{2} -r^0 \bigg)^2 - \vert \vec{r} \vert^2 \bigg)^{1+ \frac{d}{2}} + \nonumber \eeq
\beq + \frac{I_{d0}}{4} \int_{p \prec r \prec q} d^d r \; (r^1)^2 (r^2)^2 \bigg( \bigg(\frac{\tau}{2} -r^0 \bigg)^2 - \vert \vec{r} \vert^2 \bigg)^{d/2} \eeq 

{\bf Calculation 6}

\beq \int_{p \prec r \prec s \prec q} d^d r d^d s \; (r^0 s^1)(r^1 s^0) = \int_{p \prec r \prec q} \bigg(d^d r \; r^0 r^1 \int_{r \prec s \prec q} d^d s \; s^0 s^1 \bigg) \eeq
Note that 
\beq s^0 = \gamma \bigg(s^{\prime 0} + \frac{v^1}{v^0} s^{\prime 1} \bigg) + \frac{q^0+r^0}{2} = \nonumber \eeq
\beq = \frac{\frac{\tau}{2} -r^0}{\sqrt{\big(\frac{\tau}{2}-r^0 \big)^2 - \vert \vec{r} \vert^2}} \bigg(s^{\prime 0} + \frac{q^1-r^1}{\frac{\tau}{2} -r^0} s^{\prime 1} \bigg) + \frac{\frac{\tau}{2}+r^0}{2} = \nonumber \eeq 
\beq = \frac{\frac{\tau}{2}-r^0}{\sqrt{\big(\frac{\tau}{2}-r^0 \big)^2 - \vert \vec{r} \vert^2}} \bigg(s^{\prime 0} - \frac{r^1 s^{\prime 1}}{\frac{\tau}{2}-r^0} \bigg) + \frac{\frac{\tau}{2}+r^0}{2} \label{s0Lawrentz}\eeq
Therefore, by substitutting Eq \ref{skLawrentz} and \ref{s0Lawrentz} we obtain
\beq \int_{p \prec r \prec s \prec q} d^d r d^d s \; (r^0 s^1)(r^1 s^0) = \int_{p \prec r \prec q} \bigg(d^d r \; r^0 r^1 \int_{r \prec s \prec q} d^d s \; s^0 s^1 \bigg) = \nonumber \eeq
\beq = \int_{p \prec r \prec q} \Bigg(d^d r \; r^0 r^1 \int_{r \prec s \prec q} \Bigg( \Bigg( \frac{\frac{\tau}{2}-r^0}{\sqrt{\big(\frac{\tau}{2}-r^0\big)^2-\vert \vec{r} \vert^2}} \Bigg)^2 \bigg(s^{\prime 1} \bigg(- \frac{r^1s^{\prime 1}}{\frac{\tau}{2}-r^0} \bigg) - \frac{r^1s^{\prime 0}}{\frac{\tau}{2}-r^0} s^{\prime 0} \bigg) + \frac{r^1}{2} \frac{\frac{\tau}{2}+r^0}{2} \bigg) \bigg) = \nonumber  \eeq 
\beq = \int_{p \prec r \prec q} \bigg( d^d r \; r^0 r^1 \int_{r \prec s \prec q} d^d s' \bigg(- \frac{r^1 (s^{\prime 1})^2 \big(\frac{\tau}{2}-r^0 \big)}{\big(\frac{\tau}{2}-r^0 \big)^2 - \vert \vec{r} \vert^2} - \frac{r^1 (s^{\prime 0})^2 \big(\frac{\tau}{2}-r^0 \big)}{\big(\frac{\tau}{2}-r^0 \big)^2 - \vert \vec{r} \vert^2} + \frac{r^1}{2} \frac{\frac{\tau}{2}+r^0}{2} \bigg) \bigg)  \eeq 
If we now integrate out $s'$-s, we obtain 
\beq \int_{p \prec r \prec s \prec q} d^d r d^d s \; (r^0 s^1)(r^1 s^0) = \nonumber \eeq 
\beq = \int_{p \prec r \prec q} \bigg(d^d r \; r^0 r^1 \bigg(- \frac{r^1 \big(\frac{\tau}{2} -r^0 \big)}{\big(\frac{\tau}{2}-r^0\big)^2 - \vert \vec{r} \vert^2} I_{d2} \bigg(\bigg(\frac{\tau}{2}- r^0 \bigg)^2 - \vert \vec{r} \vert^2 \bigg)^{1+\frac{d}{2}}- \nonumber \eeq 
\beq - \frac{r^1 \big(\frac{\tau}{2}-r^0 \big)}{\big(\frac{\tau}{2}-r^0 \big)^2 - \vert \vec{r} \vert^2} J_{d2} \bigg( \bigg( \frac{\tau}{2} -r_0 \bigg)^2 - \vert \vec{r} \vert^2 \bigg)^{1+\frac{d}{2}} + \frac{r^1}{2} \frac{\frac{\tau}{2}+r^0}{2} I_{d0} \bigg( \bigg(\frac{\tau}{2} -r^0 \bigg)^2 - \vert \vec{r} \vert^2 \bigg)^{d/2} \bigg) \bigg) = \nonumber \eeq 
\beq = \bigg(I_{d2}+J_{d2}+ \frac{I_{d0}}{4} \bigg) \int_{p \prec r \prec q} d^d r \; (r^0)^2 (r^1)^2 \bigg( \bigg(\frac{\tau}{2}-r^0 \bigg)^2 - \vert \vec{r} \vert^2 \bigg)^{d/2} \eeq 

{\bf Adding together the above calculations} 

Let us now substitute the results of the above list of calculations into Eq \ref{SumOfDoubleIntegrals}: 
\beq  \int_{p \prec r \prec s \prec q} d^d r d^d s \; \Phi^2_{prsqp} = \nonumber \eeq
\beq = \frac{1}{2} F_{0k} F_{0k} \int d^d r d^d s \; \bigg((r^0)^2 (s^k)^2 - r^0 s^0 r^k s^k + \frac{\tau^2}{4} (r^k)^2 + \tau^2 r^k s^k \bigg) + \nonumber \eeq
\beq + \frac{1}{4} F_{ij} F_{ij} \int d^d r d^d s ((r^i)^2 (s^j)^2 - r^ir^j s^is^j) = \nonumber \eeq 
\beq = \frac{1}{2} F_{0k} F_{0k} \bigg\{ \bigg[ I_{d2} \int_{p \prec r \prec q} d^d r \; (r^0)^2 \bigg(\frac{\tau}{2}-r^0\bigg)^2 \bigg(\bigg(\frac{\tau}{2}-r^0\bigg)^2- \vert \vec{r} \vert^2 \bigg)^{d/2} + \nonumber \eeq
\beq + \bigg(\frac{I_{d0}}{4}+J_{d2}\bigg) \int_{p \prec r \prec q} d^d r \; (r^1)^2 (r^0)^2 \bigg(\bigg(\frac{\tau}{2} -r^0 \bigg)^2 - \vert \vec{r} \vert^2 \bigg)^{d/2} \bigg] + \nonumber \eeq 
\beq + \bigg[ \bigg(I_{d2}+J_{d2}+ \frac{I_{d0}}{4} \bigg) \int_{p \prec r \prec q} d^d r \; (r^0)^2 (r^1)^2 \bigg( \bigg(\frac{\tau}{2}-r^0 \bigg)^2 - \vert \vec{r} \vert^2 \bigg)^{d/2} \bigg] + \nonumber \eeq 
\beq + \frac{\tau^2}{4} \bigg[I_{d0}  \int_{p \prec r \prec q} \bigg(d^d r \; (r^1)^2  \bigg(\bigg(\frac{\tau}{2}-r^0\bigg)^2-\vert \vec{r} \vert^2 \bigg)^{d/2} \bigg) \bigg] + \nonumber \eeq 
\beq + \tau^2  \bigg[ \frac{I_{d0}}{2} \int_{p \prec r \prec q} d^d r \; \bigg( (r^1)^2 \bigg( \bigg(\frac{\tau}{2} -r^0 \bigg)^2 - \vert \vec{r} \vert^2 \bigg)^{d/2} \bigg) \bigg] \bigg\} + \nonumber \eeq 
\beq + \frac{1}{4} F_{ij} F_{ij} \bigg\{ I_{d2} \int_{p \prec r \prec q} \bigg(d^d r \; (r^1)^2 \bigg(\bigg(\frac{\tau}{2}-r^0\bigg)^2 - \vert \vec{r} \vert^2 \bigg)^{1+ \frac{d}{2} }\bigg) \bigg] - \nonumber \eeq 
\beq - \bigg[ J_{d2} \int_{p \prec r \prec q} d^d r (r^1)^2 (r^2)^2 \bigg(\bigg(\frac{\tau}{2} -r^0 \bigg)^2 - \vert \vec{r} \vert^2 \bigg)^{d/2} + \nonumber \eeq
\beq + \frac{I_{d0}}{4} \int_{p \prec r \prec q} d^d r \; (r^1)^2 (r^2)^2 \bigg( \bigg(\frac{\tau}{2} -r^0 \bigg)^2 - \vert \vec{r} \vert^2 \bigg)^{d/2} \bigg] \bigg\}  \label{FromDoubleToSingleManyTerms} \eeq
Now notice that in the last expression the second integral is the same as the third, the fourth is the same as the fifth, and the seventh is the same as the eigth. Lets therefore combine the coefficients to obtain shorter expression: 
\beq  \int_{p \prec r \prec s \prec q} d^d r d^d s \; \Phi^2_{prsqp} = \nonumber \eeq
\beq = \frac{1}{2} F_{0k} F_{0k} \bigg[ I_{d2} \int_{p \prec r \prec q} d^d r \; (r^0)^2 \bigg(\frac{\tau}{2}-r^0\bigg)^2 \bigg(\bigg(\frac{\tau}{2}-r^0\bigg)^2- \vert \vec{r} \vert^2 \bigg)^{d/2} + \nonumber \eeq
\beq + \bigg(\frac{I_{d0}}{4}+J_{d2} + I_{d2}+J_{d2}+ \frac{I_{d0}}{4} \bigg) \int_{p \prec r \prec q} d^d r \; (r^1)^2 (r^0)^2 \bigg(\bigg(\frac{\tau}{2} -r^0 \bigg)^2 - \vert \vec{r} \vert^2 \bigg)^{d/2} + \nonumber \eeq 
\beq + \bigg(\frac{\tau^2}{4}I_{d0}+ \tau^2 \frac{I_{d0}}{2} \bigg)   \int_{p \prec r \prec q} \bigg(d^d r \; (r^1)^2  \bigg(\bigg(\frac{\tau}{2}-r^0\bigg)^2-\vert \vec{r} \vert^2 \bigg)^{d/2} \bigg) \bigg] + \nonumber \eeq 
\beq + \frac{1}{4} F_{ij} F_{ij} \bigg[ I_{d2} \int_{p \prec r \prec q} \bigg(d^d r \; (r^1)^2 \bigg(\bigg(\frac{\tau}{2}-r^0\bigg)^2 - \vert \vec{r} \vert^2 \bigg)^{1+ \frac{d}{2} }\bigg)  - \nonumber \eeq 
\beq - \bigg( J_{d2}+ \frac{I_{d0}}{4} \bigg) \int_{p \prec r \prec q} d^d r (r^1)^2 (r^2)^2 \bigg(\bigg(\frac{\tau}{2} -r^0 \bigg)^2 - \vert \vec{r} \vert^2 \bigg)^{d/2} \bigg] = \nonumber \eeq
\beq = \frac{1}{2} F_{0k} F_{0k} \bigg[ I_{d2} \int_{p \prec r \prec q} d^d r \; (r^0)^2 \bigg(\frac{\tau}{2}-r^0\bigg)^2 \bigg(\bigg(\frac{\tau}{2}-r^0\bigg)^2- \vert \vec{r} \vert^2 \bigg)^{d/2} + \nonumber \eeq
\beq + \bigg(\frac{I_{d0}}{2} + I_{d2}+2J_{d2} \bigg) \int_{p \prec r \prec q} d^d r \; (r^1)^2 (r^0)^2 \bigg(\bigg(\frac{\tau}{2} -r^0 \bigg)^2 - \vert \vec{r} \vert^2 \bigg)^{d/2} + \nonumber \eeq 
\beq + \frac{3\tau^2}{4} I_{d0}   \int_{p \prec r \prec q} \bigg(d^d r \; (r^1)^2  \bigg(\bigg(\frac{\tau}{2}-r^0\bigg)^2-\vert \vec{r} \vert^2 \bigg)^{d/2} \bigg) \bigg] + \nonumber \eeq 
\beq + \frac{1}{4} F_{ij} F_{ij} \bigg[ I_{d2} \int_{p \prec r \prec q} \bigg(d^d r \; (r^1)^2 \bigg(\bigg(\frac{\tau}{2}-r^0\bigg)^2 - \vert \vec{r} \vert^2 \bigg)^{1+ \frac{d}{2} }\bigg)  - \nonumber \eeq 
\beq - \bigg( J_{d2}+ \frac{I_{d0}}{4} \bigg) \int_{p \prec r \prec q} d^d r (r^1)^2 (r^2)^2 \bigg(\bigg(\frac{\tau}{2} -r^0 \bigg)^2 - \vert \vec{r} \vert^2 \bigg)^{d/2} \bigg]  \label{FewerTerms} \eeq

where the values of the above single variable integrals will be computed in the remained of this paper( one can find those values by looking at Eq \ref{IntegralComplicatedFinal} and making appropriate substitutions). This would give us a linear combination between electric and magnetic contributions to Lagrangian and then, if we add either Eq \ref{ElectricTriangle} and/or Eq \ref{ElectricSquare} (depending on our taste) with appropriately adjusted coefficients, we would get the Lorentz covariant expression.

\subsection*{Areas of spheres and volumes of balls in $d$ dimensions}

Our final task is to evaluate the above single integrals, which is now purely math rather than physics. It is easy to see that in the process of performing the above integrals, one would most likely encounter the integrals over the spheres and, in evaluating the latter, one would encounter areas of the spheres which, in turn, are computed by means of integrals over Gaussians. Let us, therefore, proceed systematically: start out from area of the spheres and Gaussian integrals and then proceed to integral over the spheres and balls, and then finally use those integrals in evaluating integrals over Alexandrov sets. For the sake of completeness (which won't require too much extra work) we will include the derivations of some of the things that are well known (such as the integral over single variable Gaussian and so forth) and work towards what is more obscure. 

In $d$-dimensional Euclidean space, the area of the sphere is $a_d r^{d-1}$. In order to find $a_d$, we will integrate $e^{- \vert \vec{x} \vert^2/2}$ in two different ways and compare our answers. On the one hand, 
\beq \int d^d x \; e^{- \vert \vec{x} \vert^2/2} = \prod_{k=1}^d \int_{- \infty}^{\infty} dx^k e^{- x_k^2/2} = \bigg( \int_{- \infty}^{\infty} dx \; e^{-x^2/2} \bigg)^d \eeq
On the other hand,
\beq \int d^d x \; e^{- \vert \vec{x} \vert^2/2} = \int_0^{\infty} dr \; a_d r^{d-1} e^{- r^2/2} = a_d \int_0^{\infty} dr \; r^{d-1} e^{-r^2/2} \eeq
from which it follows that 
\beq a_d = \frac{\big( \int_{- \infty}^{\infty} dx \; e^{-x^2/2} \big)^d}{\int_0^{\infty} dr \; r^{d-1} e^{-r^2/2}} \label{RatioIntegrals} \eeq 
By definition of $\pi$, 
\beq a_2 = 2 \pi \eeq
and, therefore, 
\beq 2 \pi = \frac{\big( \int_{- \infty}^{\infty} dx \; e^{-x^2/2} \big)^2}{\int_0^{\infty} dr \; r^{2-1} e^{-r^2/2}} =  \frac{\big( \int_{- \infty}^{\infty} dx \; e^{-x^2/2} \big)^2}{\int_0^{\infty} dr \; r e^{-r^2/2}} \label{piDefn} \eeq
But
\beq \int_0^{\infty} dr \; re^{-r^2/2} = \int_0^{\infty} e^{-r^2/2} d \frac{r^2}{2} = -e^{-r^2/2} \bigg\vert_0^{\infty} = -(0-1) =1 \label{ExpInd1} \eeq 
and, therefore, Eq \ref{piDefn} becomes
\beq 2 \pi = \frac{\big( \int_{- \infty}^{\infty} dx \; e^{-x^2/2} \big)^2}{\int_0^{\infty} dr \; r e^{-r^2/2}} =  \bigg( \int_{- \infty}^{\infty} dx \; e^{-x^2/2} \bigg)^2 \eeq 
which implies that 
\beq \int_{- \infty}^{\infty} dx \; e^{-x^2/2} = \sqrt{2 \pi} \label{Review} \eeq 
Returning to Eq \ref{RatioIntegrals}, we now obtain
\beq a_d = \frac{\big( \int_{- \infty}^{\infty} dx \; e^{-x^2/2} \big)^d}{\int_0^{\infty} dr \; r^{d-1} e^{-r^2/2}} = \frac{(2 \pi)^{d/2}}{\int_0^{\infty} dr \; r^{d-1} e^{-r^2/2}} \label{IntegralInDenominator} \eeq
Thus, our task is to compute the integral in the denominator. The induction step goes as follows:
\beq \int_0^{\infty} x^k e^{-x^2/2} dx = \int_0^{\infty} x^{k-1} e^{-x^2/2} d \bigg( \frac{x^2}{2}\bigg) = \nonumber \eeq
\beq = - \int_0^{\infty} x^{k-1} d e^{-x^2/2}  = -x^{k-1} e^{-x^2/2} \bigg\vert_0^{\infty} + \int_0^{\infty} e^{-x^2/2} dx^{k-1} = \nonumber \eeq
\beq= \int_0^{\infty} e^{-x^2/2} (k-1)x^{k-2} dx  = (k-1) \int_0^{\infty} e^{-x^2/2} x^{k-2} dx \eeq 
Now lets do separate calculations for even and odd powers. In case of odd power, 
\beq \int_0^{\infty} x^{2n+1} e^{-x^2/2} dx = (2n) \int_0^{\infty} x^{2n-1} e^{-x^2/2} dx = \nonumber \eeq
\beq = (2n)(2n-2) \int_0^{\infty} x^{2n-3}  e^{-x^2/2} dx = \bigg(\prod_{k=1}^n (2k) \bigg) \int_0^{\infty} xe^{-x^2/2}dx= \nonumber \eeq
 \beq =  2^n n! \int_0^{\infty} xe^{-x^2/2}dx =^{Eq \; \ref{ExpInd1}} 2^n n! \eeq 
On the other hand, in case of even power, 
\beq \int_0^{\infty} x^{2n}  e^{-x^2/2} dx = (2n-1) \int_0^{\infty} x^{2n-2}  e^{-x^2/2} dx = \nonumber \eeq
\beq = (2n-1)(2n-3) \int_0^{\infty} x^{2n-4}  e^{-x^2/2} dx = \cdots = \bigg( \prod_{k=1}^n (2k-1) \bigg) \int_0^{\infty} e^{-x^2/2} dx = \nonumber \eeq 
\beq = \bigg( \prod_{k=1}^n (2k-1) \bigg) \frac{\int_{- \infty}^{\infty} e^{-x^2/2}}{2} dx =^{Eq \; \ref{Review}} \bigg( \prod_{k=1}^n (2k-1) \bigg) \frac{\sqrt{2 \pi}}{2} = \sqrt{\frac{\pi}{2}}  \prod_{k=1}^n (2k-1) = \nonumber \eeq
\beq = \frac{\prod_{k=1}^n (2k)}{\prod_{l=1}^{2n} l}  \sqrt{\frac{\pi}{2}}=  \frac{2^n \prod_{k=1}^n k}{\prod_{l=1}^{2n} l}  \sqrt{\frac{\pi}{2}}=  \frac{2^n n!}{(2n)!}  \sqrt{\frac{\pi}{2}}\eeq 
We can combine the answers for even and odd powers in the following way: 
\beq \int x^m  e^{-x^2/2} dx = \frac{2^{\lfloor m/2 \rfloor}\lfloor \frac{m}{2} \rfloor !}{(m!)^{(1+(-1)^m)/2}} \bigg(\frac{\pi}{2} \bigg)^{(1+(-1)^m)/4} \eeq 
By substituting this into Eq \ref{IntegralInDenominator} we obtain
\beq a_d = \frac{(2 \pi)^{d/2}}{\int_0^{\infty} dr \; r^{d-1} e^{-r^2/2}} = \frac{(2 \pi)^{d/2}((d-1)!)^{(1+(-1)^{d-1})/2}}{2^{\lfloor (d-1)/2 \rfloor}\lfloor \frac{d-1}{2} \rfloor !} \bigg(\frac{2}{\pi} \bigg)^{(1+(-1)^{d-1})/4} = \nonumber \eeq
\beq = \frac{2^{\frac{d}{2}- \lfloor \frac{d-1}{2} \rfloor + \frac{1+(-1)^{d-1}}{4}} \pi^{\frac{d}{2}- \frac{1+(-1)^{d-1}}{4}} ((d-1)!)^{(1+(-1)^{d-1})/2}}{\lfloor \frac{d-1}{2} \rfloor!} \eeq 
Now note the following: 
\beq d=2k \Longrightarrow \frac{d}{2} - \bigg\lfloor \frac{d-1}{2} \bigg\rfloor = k - \bigg\lfloor k- \frac{1}{2} \bigg\rfloor = k - (k-1) = 1 \eeq 
\beq d=2k+1 \Longrightarrow  \frac{d}{2} - \bigg\lfloor \frac{d-1}{2} \bigg\rfloor =  k+ \frac{1}{2}- \lfloor k \rfloor =  k + \frac{1}{2} -k = \frac{1}{2} \eeq 
Thus, we can generalize it as 
\beq \frac{d}{2} - \bigg\lfloor \frac{d-1}{2} \bigg\rfloor= \frac{3+(-1)^d}{4} \eeq 
Thus, we obtain
\beq a_d =  \frac{2^{\frac{d}{2}- \lfloor \frac{d-1}{2} \rfloor + \frac{1+(-1)^{d-1}}{4}} \pi^{\frac{d}{2}- \frac{1+(-1)^{d-1}}{4}} ((d-1)!)^{(1+(-1)^{d-1})/2}}{\lfloor \frac{d-1}{2} \rfloor!} = \nonumber \eeq 
\beq =  \frac{2^{\frac{3+(-1)^d}{4} + \frac{1+(-1)^{d-1}}{4}} \pi^{\frac{d}{2}- \frac{1+(-1)^{d-1}}{4}} ((d-1)!)^{(1+(-1)^{d-1})/2}}{\lfloor \frac{d-1}{2} \rfloor!} \eeq 
Now noticing that 
\beq \frac{3+(-1)^d}{4} + \frac{1+(-1)^{d-1}}{4} = \frac{3+1}{4} + \frac{(-1)^d + (-1)^{d-1}}{4} = 1+0 = 1 \eeq 
we obtain 
\beq a_d=  \frac{2 \big(\pi^{\frac{d}{2}- \frac{1+(-1)^{d-1}}{4}}\big) ((d-1)!)^{(1+(-1)^{d-1})/2}}{\lfloor \frac{d-1}{2} \rfloor!} \eeq 
Lets check it by comparing to known answers. 
\beq a_2=  \frac{2 \big(\pi^{\frac{2}{2}- \frac{1+(-1)^{2-1}}{4}}\big) ((2-1)!)^{(1+(-1)^{2-1})/2}}{\lfloor \frac{2-1}{2} \rfloor!} = \frac{2 \pi^1 1^0}{0!} = 2 \pi \eeq 
\beq a_3=  \frac{2 \big(\pi^{\frac{3}{2}- \frac{1+(-1)^{3-1}}{4}}\big) ((3-1)!)^{(1+(-1)^{3-1})/2}}{\lfloor \frac{3-1}{2} \rfloor!} = \frac{2 \pi^1 2^1}{1!} = 4 \pi \eeq 
Finally the volume is given by 
\beq V = v_d r^d \eeq
where $v_d$ is volume of a ball of radius $1$, given by 
\beq v_d = \int_0^1 a_d r^{d-1} dr = \frac{a_dr^d}{d} \bigg\vert_0^1 = \frac{a_d}{d} =  \frac{2 \big(\pi^{\frac{d}{2}- \frac{1+(-1)^{d-1}}{4}}\big) ((d-1)!)^{(1+(-1)^{d-1})/2}}{\lfloor \frac{d-1}{2} \rfloor!d} \eeq
From the fact that $v_d = a_d/d$ its trivial that if $a_d$ work for dimensions $2$ and $3$, then $v_d$ would also work for those dimensions:
\beq d=2 \Longrightarrow v_2 = \frac{a_2}{2} = \frac{2 \pi}{2} = \pi \eeq
\beq d= 3 \Longrightarrow v_3 = \frac{a_3}{2} = \frac{4 \pi}{3} = \frac{4}{3} \pi \eeq
Since $v_d = a_d/d$ is the only thing we used in obtaining $v_d$, we don't have to check it. 

\subsection*{General single variable integral involving Gaussian}

In the previous section we have found that 
\beq a_d=  \frac{2 \big(\pi^{\frac{d}{2}- \frac{1+(-1)^{d-1}}{4}}\big) ((d-1)!)^{(1+(-1)^{d-1})/2}}{\lfloor \frac{d-1}{2} \rfloor!} \eeq 
combining this with Eq \ref{IntegralInDenominator},
\beq a_d  = \frac{(2 \pi)^{d/2}}{\int_0^{\infty} dr \; r^{d-1} e^{-r^2/2}} \eeq
we obtain
\beq \int_0^{\infty} dr \; r^{d-1} e^{-r^2/2} =  \frac{\lfloor \frac{d-1}{2} \rfloor!(2 \pi)^{d/2}}{2 \big(\pi^{\frac{d}{2}- \frac{1+(-1)^{d-1}}{4}}\big) ((d-1)!)^{(1+(-1)^{d-1})/2}} \eeq 
Note that the above is a single variable integral. Geometrically we know that $r$ is a radius; but algebraically we had to compute the single variable integral. Therefore, we might as well replace $r$ with $x$ in order to use this result in other contexts. In this case, $d$ will no longer have geometric significance, it will be just a number. Therefore, we can replace $d-1$ with $n$ (or, equivalently, $d$ with $n+1$). Making those replacements, we obtain 
\beq \int_0^{\infty} dx \; x^n e^{-x^2/2} =  \frac{\lfloor \frac{n}{2} \rfloor!(2 \pi)^{(n+1)/2}}{2 \big(\pi^{\frac{n+1}{2}- \frac{1+(-1)^n}{4}}\big) (n!)^{(1+(-1)^n)/2}} \eeq 
 Now if we are to integrate it over $(- \infty, \infty)$ instead of $[0, \infty)$, then we will get extra factor of $2$ if $n$ is even and we will have $0$ if $n$ is odd. These two statements are equivalent to the statement that we have an extra factor of $1+(-1)^n$. Thus, 
\beq \int_{- \infty}^{\infty} dx \; x^n e^{-x^2/2} =  \frac{(1+(-1)^n)\lfloor \frac{n}{2} \rfloor!(2 \pi)^{(n+1)/2}}{2 \big(\pi^{\frac{n+1}{2}- \frac{1+(-1)^n}{4}}\big) (n!)^{(1+(-1)^n)/2}} \label{1DGaussianGeneral} \eeq 
To check it, 
\beq \int_{- \infty}^{\infty} e^{-x^2/2} dx =  \frac{(1+(-1)^0)\lfloor \frac{0}{2} \rfloor!(2 \pi)^{(0+1)/2}}{2 \big(\pi^{\frac{0+1}{2}- \frac{1+(-1)^0}{4}}\big) (0!)^{(1+(-1)^0)/2}} = \frac{2*0!*(2 \pi)^{1/2}}{2 \pi^{0} (0!)^1} = \frac{2 \sqrt{2 \pi}}{2} = \sqrt{2 \pi} \eeq 
 
\subsection*{Integrals over a sphere}

Let $S (r)$ be a sphere of radius $r$ in $d$ dimensions around the origin, and let $S_{p_1 \cdots p_d}$ be an integral of the form
\beq S_{p_1 \cdots p_d} = \int_{S(1)} dA \; (x^1)^{p_1} \cdots (x^d)^{p_d} \label{DefnOfS} \eeq 
We will use similar tric to before, where we try to integrate Gaussian. On the one hand, 
\beq \int d^d x \; e^{- \vert \vec{x} \vert^2/2} (x^1)^{p_1} \cdots (x^d)^{p_d} = \prod_{k=1}^d \int e^{-x_k^2/2} (x^k)^{p_k} dx_k \label{ProductDifferent} \eeq 
On the other hand, 
\beq \int d^d x e^{- \vert \vec{x} \vert^2/2} (x^1)^{p_1} \cdots (x^d)^{p_d} = \int_0^{\infty} \bigg(dr e^{-r^2/2} \int_{S(r)} dA \;  (x^1)^{p_1} \cdots (x^d)^{p_d} \bigg) \label{ExpandingSphere} \eeq 
From dimensional analysis we know that 
\beq \int_{S(r)} dA \; (x^1)^{p_1} \cdots (x^d)^{p_d} = r^{d-1+p_1 + \cdots + p_d} S_{p_1 \cdots p_d} \label{SDimAnalysis}\eeq 
and, therefore, Eq \ref{ExpandingSphere} becomes
\beq \int d^d x e^{- \vert \vec{x} \vert^2/2} (x^1)^{p_1} \cdots (x^d)^{p_d} = \int_0^{\infty} \bigg(dr \; e^{-r^2/2}  r^{d-1+p_1 + \cdots + p_d} S_{p_1 \cdots p_d}   \bigg) = \nonumber \eeq 
\beq = S_{p_1 \cdots p_d}\int_0^{\infty} dr \; e^{-r^2/2}  r^{d-1+p_1 + \cdots + p_d} \eeq 
By comparing this to Eq \ref{ProductDifferent} we obtain 
\beq S_{p_1 \cdots p_d}\int_0^{\infty} dr \; e^{-r^2/2}  r^{d-1+p_1 + \cdots + p_d} = \prod_{k=1}^d \int e^{-x_k^2/2} (x^k)^{p_k} dx_k \eeq
and, therefore, 
\beq S_{p_1 \cdots p_d} = \frac{\prod_{k=1}^d \int e^{-x_k^2/2} (x^k)^{p_k} dx_k}{\int_0^{\infty} dr \; e^{-r^2/2}  r^{d-1+p_1 + \cdots + p_d} } \eeq
which then, by using Eq \ref{DefnOfS} can be rewritten as 
\beq \int_{S(1)} dA \; (x^1)^{p_1} \cdots (x^d)^{p_d} = \frac{\prod_{k=1}^d \int e^{-x_k^2/2} (x^k)^{p_k} dx_k}{\int_0^{\infty} dr \; e^{-r^2/2}  r^{d-1+p_1 + \cdots + p_d} } \eeq
By using Eq \ref{1DGaussianGeneral} we can rewrite it as
\beq \int_{S(1)} dA \; (x^1)^{p_1} \cdots (x^d)^{p_d} = \nonumber \eeq
 \beq = \frac{\prod_{k=1}^d\frac{(1+(-1)^{p_k})\lfloor \frac{p_k}{2} \rfloor!(2 \pi)^{(p_k+1)/2}}{2 \big(\pi^{\frac{p_k+1}{2}- \frac{1+(-1)^{p_k}}{4}}\big) (p_k!)^{(1+(-1)^{p_k})/2}}}{\frac{(1+(-1)^{d-1+p_1+\cdots+p_d})\lfloor \frac{d-1+p_1+\cdots+p_d}{2} \rfloor!(2 \pi)^{({d-1+p_1+\cdots+p_d}+1)/2}}{2 \big(\pi^{\frac{d-1+p_1+\cdots+p_d+1}{2}- \frac{1+(-1)^{d-1+p_1+\cdots+p_d}}{4}}\big) (({d-1+p_1+\cdots+p_d})!)^{(1+(-1)^{d-1+p_1+\cdots+p_d})/2}} } = \nonumber \eeq
\beq = \frac{2 \big(\pi^{\frac{d-1+p_1+\cdots+p_d+1}{2}- \frac{1+(-1)^{d-1+p_1+\cdots+p_d}}{4}}\big) (({d-1+p_1+\cdots+p_d})!)^{(1+(-1)^{d-1+p_1+\cdots+p_d})/2}}{(1+(-1)^{d-1+p_1+\cdots+p_d})\lfloor \frac{d-1+p_1+\cdots+p_d}{2} \rfloor!(2 \pi)^{({d-1+p_1+\cdots+p_d}+1)/2}} \times \nonumber \eeq
\beq \times \prod_{k=1}^d\frac{(1+(-1)^{p_k})\lfloor \frac{p_k}{2} \rfloor!(2 \pi)^{(p_k+1)/2}}{2 \big(\pi^{\frac{p_k+1}{2}- \frac{1+(-1)^{p_k}}{4}}\big) (p_k!)^{(1+(-1)^{p_k})/2}} = S_{p_1 \cdots p_d} \label{SFinal}\eeq 

\subsection*{Integrals over Alexandrov set}

Consider an Alexandrov set $\alpha (p,q)$ and choose coordinate system so that $t$-axis passes through $p$ and $q$ with origin in the middle. Furthermore, suppose the distance between $p$ and $q$ is $\tau$. Finally, we have $d-1$ space dimensions and $1$ time dimension.  Therefore, in this coordinate system, 
\beq p = \bigg(- \frac{\tau}{2}, 0, \cdots, 0 \bigg) \; , \; q = \bigg(\frac{\tau}{2},0, \cdots, 0 \bigg) \eeq 
Suppose we are interested in the integral of the form 
\beq \int_{\alpha (p,q)} d^d r \; \bigg(\bigg(\frac{\tau}{2} -r^0 \bigg)^2 - \vert \vec{r} \vert^2 \bigg)^{p/2} \bigg(\frac{\tau}{2} -x^0 \bigg)^{p_0} (x^1)^{p_1} \cdots (x^{d-1})^{p_{d-1}} \eeq
where the reason we chose to write $(\tau/2 - x^0)^{p_0}$ instead of $(x^0)^{p_0}$ is simply that the answer is simpler in this form, as we will find out. Clearly, the term in the brackets is the distance from $r$ to $q$ taken to the power $p$. We will slice the volume into slices, so that in each slice the distance from $r$ to $q$ is $\xi$, and then integrate over $\xi$ ranging from $0$ to $\tau$. Now, on any given $\xi= \rm const$ hypersurface, $\vec{r}$ and $t$ behave as 
\beq t = \frac{\tau}{2} - \xi \cosh \eta \; , \; \vert \vec{r} \vert = \xi \sinh \eta \eeq 
where $\eta$ ranges from $0$ to $\eta_0$. Now, $\eta= \eta_0$ corresponds to the intersection of the above hypersurface with the lightcone of $p$. In other words, at $\eta= \eta_0$ the distance to $p$ is $0$. The general expression for distance to $p$ is given by 
\beq \big((\tau- \xi \cosh \eta)^2 - (\xi \sinh \eta)^2 \big)^{1/2} = \big(\tau^2 - 2 \tau \xi \cosh \eta + \xi^2 \cosh^2 \eta - \xi^2 \sinh^2 \eta \big)^{1/2} = \nonumber \eeq 
\beq = \big(\tau^2 - 2 \tau \xi \cosh \eta + \xi^2 (\cosh^2 \eta - \sinh^2 \eta) \big)^{1/2} = \big(\tau^2 - 2 \tau \xi \cosh \eta + \xi^2 \big)^{1/2} \eeq
Therefore, we can find $\eta_0$ by using 
\beq \big(\tau^2 - 2 \tau \xi \cosh \eta_0 + \xi^2 \big)^{1/2} = 0 \Longrightarrow 2 \tau \xi \cosh \eta_0 = \tau^2 + \xi^2 \Longrightarrow \nonumber \eeq 
\beq \Longrightarrow \cosh \eta_0 = \frac{\tau^2 + \xi^2}{2 \tau \xi} = \frac{\frac{\tau^2}{\tau \xi} + \frac{\xi^2}{\tau \xi}}{2} = \frac{\frac{\tau}{\xi} + \frac{\xi}{\tau}}{2} \eeq 
Since we also know that 
\beq \cosh \eta_0 = \frac{e^{\eta_0} + e^{- \eta_0}}{2} \eeq
we have 
\beq  \frac{e^{\eta_0} + e^{- \eta_0}}{2} = \frac{\frac{\tau}{\xi} + \frac{\xi}{\tau}}{2}  \eeq
If we multiply it by $e^{\eta_0}$, we would have quadratic in $e^{\eta_0}$. Thus, we know that we can have at most two solutions. And we can read off what those two solutions are:  either $e^{\eta_0}= \tau/\xi$ and, therefore, $e^{-\eta_0}= \xi/\tau$, or else $e^{\eta_0}= \xi/\tau$ and, therefore, $e^{-\eta_0}= \tau/\xi$. Since $0 < \xi < \tau$ and we choose a convention where $\eta>0$, we stick to the choice 
\beq e^{\eta_0} = \frac{\tau}{\xi} \Longrightarrow \eta_0 = \ln \frac{\tau}{\xi} \eeq 
Now by means of Eq \ref{SDimAnalysis} while remembering to use $d-1$ in place of $d$ (since $d$ is the number of spacetime dimensions hence we have only $d-1$ space dimensions) and also remembering to use 
\beq dr = \cosh \eta \; d \eta \eeq
instead of $d \eta$, we can perform the integral in the following way:
\beq \int_{\alpha (p,q)} d^d r \; \bigg(\bigg(\frac{\tau}{2} -r^0 \bigg)^2 - \vert \vec{r} \vert^2 \bigg)^{p/2} \bigg(\frac{\tau}{2} -x^0 \bigg)^{p_0} (x^1)^{p_1} \cdots (x^{d-1})^{p_{d-1}} = \nonumber \eeq
\beq = \int_0^{\tau} \bigg(d \xi \int_0^{\ln \frac{\tau}{\xi}} \xi^p (\xi \sinh \eta)^{d-2+p_1 + \cdots + p_{d-1}} (\xi \cosh \eta)^{p_0} (\xi \cosh \eta \; d \eta) \bigg) \eeq
Combining the powers of $\xi$,
\beq p+ (d-2+p_1+\cdots+p_{d-1})+p_0+1 = p+p_0+p_1+ \cdots + p_{d-1}+d-2+1 = \nonumber \eeq
\beq = p+p_0 + \cdots +p_{d-1}+d-1 \eeq 
we obtain 
\beq \int_{\alpha (p,q)} d^d r \; \bigg(\bigg(\frac{\tau}{2} -r^0 \bigg)^2 - \vert \vec{r} \vert^2 \bigg)^{p/2} \bigg(\frac{\tau}{2} -x^0 \bigg)^{p_0} (x^1)^{p_1} \cdots (x^{d-1})^{p_{d-1}} = \nonumber \eeq
\beq =  S_{p_1 \cdots p_{d-1}} \int_0^{\tau} \bigg(d \xi \; \xi^{ p+p_0 + \cdots +p_{d-1}+d-1} \int_0^{\ln \frac{\tau}{\xi}} (\sinh \eta)^{d-2+p_1+ \cdots +p_{d-1}} (\cosh \eta)^{p_0+1} d \eta \bigg) = \nonumber \eeq 
\beq =  S_{p_1 \cdots p_{d-1}} \int_0^{\tau} \bigg(d \xi \; \xi^{ p+p_0 + \cdots +p_{d-1}+d-1} \int_0^{\ln \frac{\tau}{\xi}} \bigg(\frac{e^{\eta}-e^{-\eta}}{2} \bigg)^{d-2+p_1+ \cdots +p_{d-1}} \bigg(\frac{e^{\eta}+e^{-\eta}}{2} \bigg)^{p_0+1} d \eta \bigg)  \eeq 
Noticing that 
\beq \frac{1}{2^{d-2+p_1+ \cdots +p_{d-1}}} \frac{1}{2^{p_0+1}}= \frac{1}{2^{(d-2+p_1+ \cdots +p_{d-1})+(p_0+1)}} = \frac{1}{2^{d-1+p_0 + \cdots +p_{d-1}}} \eeq 
and also that 
\beq (e^{\eta}+e^{-\eta})^{d-2+p_1+ \cdots + p_{d-1}} (e^{\eta}+e^{-\eta})^{p_0+1} = \nonumber \eeq
 \beq = \bigg(\sum_{k=0}^{d-2+p_1+ \cdots + p_{d-1}} {d-2+p_1+ \cdots + p_{d-1} \choose k} (e^{\eta})^k (-e^{- \eta})^{d-2+p_1+ \cdots + p_{d-1}} \bigg)   \times \nonumber \eeq
 \beq \times \bigg( \sum_{l=0}^{p_0+1} {p_0+1 \choose l} (e^{\eta})^l (e^{- \eta})^{p_0+1-l} \bigg) = \nonumber \eeq 
\beq = \sum_{k=0}^{d-2+p_1+ \cdots + p_{d-1}} \; \; \; \sum_{l=0}^{p_0+1} \bigg[ {d-2+p_1+ \cdots + p_{d-1} \choose k} {p_0+1 \choose l} \times \nonumber \eeq
\beq \times (-1)^{d-2+p_1+ \cdots + p_{d-1}} e^{\eta (k+(d-2+p_1+ \cdots + p_{d-1})+l+ (p_0+1-l))} \bigg] = \nonumber \eeq 
\beq = \sum_{k=0}^{d-2+p_1+ \cdots + p_{d-1}} \; \; \; \sum_{l=0}^{p_0+1} \bigg[ {d-2+p_1+ \cdots + p_{d-1} \choose k} {p_0+1 \choose l} \times \nonumber \eeq
\beq \times (-1)^{d-2+p_1+ \cdots + p_{d-1}} e^{\eta (k+d-1+p_0+ \cdots + p_{d-1})} \bigg]  \eeq 
we obtain 
\beq \int_{\alpha (p,q)} d^d r \; \bigg(\bigg(\frac{\tau}{2} -r^0 \bigg)^2 - \vert \vec{r} \vert^2 \bigg)^{p/2} \bigg(\frac{\tau}{2} -x^0 \bigg)^{p_0} (x^1)^{p_1} \cdots (x^{d-1})^{p_{d-1}} = \nonumber \eeq
\beq =  S_{p_1 \cdots p_{d-1}} \int_0^{\tau} \bigg(d \xi \; \xi^{ p+p_0 + \cdots +p_{d-1}+d-1} \int_0^{\ln \frac{\tau}{\xi}} \bigg(\frac{e^{\eta}-e^{-\eta}}{2} \bigg)^{d-2+p_1+ \cdots +p_{d-1}} \bigg(\frac{e^{\eta}+e^{-\eta}}{2} \bigg)^{p_0+1} d \eta \bigg)  = \nonumber \eeq 
\beq =  \frac{S_{p_1 \cdots p_{d-1}}}{2^{d-1+p_0+\cdots+p_{d-1}}} \int_0^{\tau} \bigg \{ d \xi \; \xi^{ p+p_0 + \cdots +p_{d-1}+d-1} \times \nonumber \eeq
\beq \times \int_0^{\ln \frac{\tau}{\xi}} \bigg[ d \eta \sum_{k=0}^{d-2+p_1+ \cdots + p_{d-1}} \; \; \; \sum_{l=0}^{p_0+1} \bigg[ {d-2+p_1+ \cdots + p_{d-1} \choose k} {p_0+1 \choose l} \times \nonumber \eeq
\beq \times (-1)^{d-2+p_1+ \cdots + p_{d-1}} e^{\eta (k+d-1+p_0+ \cdots + p_{d-1})} \bigg] \bigg] \bigg \}= \nonumber \eeq 
\beq =  \frac{S_{p_1 \cdots p_{d-1}}}{2^{d-1+p_0+\cdots+p_{d-1}}} \int_0^{\tau} \bigg \{ d \xi \; \xi^{ p+p_0 + \cdots +p_{d-1}+d-1} \times \nonumber \eeq
\beq  \times \sum_{k=0}^{d-2+p_1+ \cdots + p_{d-1}} \; \; \; \sum_{l=0}^{p_0+1} \bigg[ {d-2+p_1+ \cdots + p_{d-1} \choose k} {p_0+1 \choose l} \times \nonumber \eeq
\beq \times (-1)^{d-2+p_1+ \cdots + p_{d-1}} \frac{e^{\eta (k+d-1+p_0+ \cdots + p_{d-1})}}{(k+d-1+p_0+ \cdots + p_{d-1}} \bigg] \bigg\vert_{\eta=0}^{\eta=\ln \frac{\tau}{\xi}} \bigg \}\eeq 
Noticing that 
\beq \eta = \ln \frac{\tau}{\xi} \Longrightarrow e^{\eta} = \frac{\tau}{\xi} \Longrightarrow e^{\eta (k+d-1+p_0+\cdots+p_{d-1})} = \bigg(\frac{\tau}{\xi} \bigg)^{k+d-1+p_0+ \cdots +p_{d-1}} \eeq 
and also that 
\beq \eta=0 \Longrightarrow e^{\eta (k+d-1+p_0+\cdots+p_{d-1})} =e^0 =1 \eeq
we obtain 
\beq \int_{\alpha (p,q)} d^d r \; \bigg(\bigg(\frac{\tau}{2} -r^0 \bigg)^2 - \vert \vec{r} \vert^2 \bigg)^{p/2} \bigg(\frac{\tau}{2} -x^0 \bigg)^{p_0} (x^1)^{p_1} \cdots (x^{d-1})^{p_{d-1}} = \nonumber \eeq
\beq  = \frac{S_{p_1 \cdots p_{d-1}}}{2^{d-1+p_0+\cdots+p_{d-1}}} \int_0^{\tau} \bigg \{ d \xi \; \xi^{ p+p_0 + \cdots +p_{d-1}+d-1} \times \nonumber \eeq
\beq  \times \sum_{k=0}^{d-2+p_1+ \cdots + p_{d-1}} \; \; \; \sum_{l=0}^{p_0+1} \bigg[ {d-2+p_1+ \cdots + p_{d-1} \choose k} {p_0+1 \choose l} \times \nonumber \eeq
\beq \times (-1)^{d-2+p_1+ \cdots + p_{d-1}} \frac{(\frac{\tau}{\xi})^{k+d-1+p_0+ \cdots + p_{d-1}}-1}{(k+d-1+p_0+ \cdots + p_{d-1}} \bigg]  \bigg \} \eeq 
Now we note that, on the one hand, there is a power of $\xi$ up front and, on the other hand, there is a sum involving power of $\xi$ at the end. We now multiply the former by the latter: 
\beq  \xi^{ p+p_0 + \cdots +p_{d-1}+d-1} \bigg( \bigg(\frac{\tau}{\xi} \bigg)^{k+d-1+p_0+ \cdots + p_{d-1}}-1 \bigg) = \tau^{k+d-1+p_0+ \cdots + p_{d-1}} \xi^{p-k}  - \xi^{p+p_0 + \cdots +p_{d-1}+d-1} \eeq 
Thus, we write 
\beq \int_{\alpha (p,q)} d^d r \; \bigg(\bigg(\frac{\tau}{2} -r^0 \bigg)^2 - \vert \vec{r} \vert^2 \bigg)^{p/2} \bigg(\frac{\tau}{2} -x^0 \bigg)^{p_0} (x^1)^{p_1} \cdots (x^{d-1})^{p_{d-1}} = \nonumber \eeq
\beq  = \frac{S_{p_1 \cdots p_{d-1}}}{2^{d-1+p_0+\cdots+p_{d-1}}} \int_0^{\tau} \bigg \{ d \xi \;  \sum_{k=0}^{d-2+p_1+ \cdots + p_{d-1}} \; \; \; \sum_{l=0}^{p_0+1} \bigg[ {d-2+p_1+ \cdots + p_{d-1} \choose k} {p_0+1 \choose l} \times \nonumber \eeq
\beq \times (-1)^{d-2+p_1+ \cdots + p_{d-1}} \frac{\tau^{k+d-1+p_0+ \cdots + p_{d-1}} \xi^{p-k}  - \xi^{p+p_0 + \cdots +p_{d-1}+d-1} }{(k+d-1+p_0+ \cdots + p_{d-1}} \bigg]  \bigg \} = \nonumber \eeq 
\beq  = \frac{S_{p_1 \cdots p_{d-1}}}{2^{d-1+p_0+\cdots+p_{d-1}}}   \sum_{k=0}^{d-2+p_1+ \cdots + p_{d-1}} \; \; \; \sum_{l=0}^{p_0+1} \bigg( {d-2+p_1+ \cdots + p_{d-1} \choose k} {p_0+1 \choose l} \times \nonumber \eeq
\beq \times (-1)^{d-2+p_1+ \cdots + p_{d-1}} \frac{\tau^{k+d-1+p_0+ \cdots + p_{d-1}} \frac{\xi^{p-k+1}}{p-k+1}  - \frac{\xi^{d+p+p_0+ \cdots + p_{d-1}}}{d+p+p_0 + \cdots + p_{d-1}}}{(k+d-1+p_0+ \cdots + p_{d-1}} \bigg) \bigg\vert^{\xi=\tau}_{\xi=0} = \nonumber \eeq 
\beq  = \frac{S_{p_1 \cdots p_{d-1}}}{2^{d-1+p_0+\cdots+p_{d-1}}}   \sum_{k=0}^{d-2+p_1+ \cdots + p_{d-1}} \; \; \; \sum_{l=0}^{p_0+1} \bigg( {d-2+p_1+ \cdots + p_{d-1} \choose k} {p_0+1 \choose l} \times \nonumber \eeq
\beq \times (-1)^{d-2+p_1+ \cdots + p_{d-1}} \frac{\tau^{k+d-1+p_0+ \cdots + p_{d-1}} \frac{\tau^{p-k+1}}{p-k+1}  - \frac{\tau^{d+p+p_0 + \cdots + p_{d-1}}}{d+p+p_0 + \cdots + p_{d-1}}}{(k+d-1+p_0+ \cdots + p_{d-1}} \bigg) = \nonumber \eeq 
\beq  = \frac{S_{p_1 \cdots p_{d-1}}}{2^{d-1+p_0+\cdots+p_{d-1}}}   \sum_{k=0}^{d-2+p_1+ \cdots + p_{d-1}} \; \; \; \sum_{l=0}^{p_0+1} \bigg( {d-2+p_1+ \cdots + p_{d-1} \choose k} {p_0+1 \choose l} \times \nonumber \eeq
\beq \times (-1)^{d-2+p_1+ \cdots + p_{d-1}} \frac{ \frac{\tau^{d+ p+p_0+\cdots + p_{d-1}}}{p-k+1}  - \frac{\tau^{d+p+p_0 + \cdots + p_{d-1}}}{d+p+p_0 + \cdots + p_{d-1}}}{(k+d-1+p_0+ \cdots + p_{d-1}} \bigg) = \nonumber \eeq
\beq  = \frac{ (-1)^{d-2+p_1+ \cdots + p_{d-1}} S_{p_1 \cdots p_{d-1}}\tau^{d+p+p_0 + \cdots + p_{d-1}}}{2^{d-1+p_0+\cdots+p_{d-1}}}  \times  \label{c=1} \eeq
\beq \times  \sum_{k=0}^{d-2+p_1+ \cdots + p_{d-1}} \; \; \; \sum_{l=0}^{p_0+1} \bigg( {d-2+p_1+ \cdots + p_{d-1} \choose k} {p_0+1 \choose l} \frac{ \frac{1}{p-k+1}  - \frac{1}{d+p+p_0 + \cdots + p_{d-1}}}{(k+d-1+p_0+ \cdots + p_{d-1}} \bigg)  \nonumber  \eeq
Finally, suppose that we replace $(\tau/2 - x^0)^{p_0}$ with $(c_1 \tau/2-x^0)^{p_{10}}(c_2 \tau/2-x^0)^{p_{20}}$. Then we use
\beq \bigg(\frac{c \tau}{2}-r^0 \bigg)^{p_0} = \bigg(\bigg(\frac{c \tau}{2}- \frac{\tau}{2} \bigg) + \bigg(\frac{\tau}{2} -r^0 \bigg) \bigg)^{p_0} = \sum_{j=1}^{p_0} {p_0 \choose j} \bigg(\frac{c \tau}{2}- \frac{\tau}{2} \bigg)^{p_0-j} \bigg(\frac{\tau}{2} -r^0 \bigg)^j  = \nonumber \eeq 
\beq = \sum_{j=1}^{p_0} {p_0 \choose j} (c-1)^{p_0-j} \bigg(\frac{\tau}{2} \bigg)^{p_0-j} \bigg(\frac{\tau}{2} -r_0 \bigg)^j \eeq 
and, therefore, 
\beq (c_1 \tau/2-x^0)^{p_{10}}(c_2 \tau/2-x^0)^{p_{20}} = \nonumber \eeq
\beq = \bigg(\sum_{j_1=1}^{p_{01}} {p_{01} \choose j_1} (c_1-1)^{p_{01}-j_1} \bigg(\frac{\tau}{2} \bigg)^{p_{01}-j_1} \bigg(\frac{\tau}{2} -r^0 \bigg)^{j_1}  \bigg) \bigg(\sum_{j_2=1}^{p_{02}} {p_{02} \choose j_2} (c_2-1)^{p_{02}-j} \bigg(\frac{\tau}{2} \bigg)^{p_{02}-j_2} \bigg(\frac{\tau}{2} -r^0 \bigg)^{j_2}  \bigg) = \nonumber \eeq 
\beq = \sum_{j_1=1}^{p_{01}} \sum_{j_2=1}^{p_{02}}  {p_{01} \choose j_1} {p_{02} \choose j_2} (c_1-1)^{p_{01}-j_1} (c_2-1)^{p_{02}-j_2} \bigg(\frac{\tau}{2} \bigg)^{p_{01}+p_{02}-j_1-j_2} \bigg(\frac{\tau}{2} -r^0 \bigg)^{j_1+j_2} \bigg) \eeq

thus obtaining 
\beq \int_{\alpha (p,q)} d^d r \; \bigg(\bigg(\frac{\tau}{2} -r^0 \bigg)^2 - \vert \vec{r} \vert^2 \bigg)^{p_0/2}  \bigg(\frac{c_1\tau}{2} -x^0 \bigg)^{p_{01}} \bigg(\frac{c_2\tau}{2} -x^0 \bigg)^{p_{02}} (x^1)^{p_1} \cdots (x^{d-1})^{p_{d-1}} =  \nonumber \eeq
\beq =  \sum_{j_1=1}^{p_{01}} \sum_{j_2=1}^{p_{02}}\bigg( {p_{01} \choose j_1} {p_{02} \choose j_2} (c_1-1)^{p_{01}-j_1}(c_2-1)^{p_{02}-j_2} \bigg(\frac{\tau}{2} \bigg)^{p_{01}+p_{02}-j_1-j_2} \times \nonumber \eeq
\beq \times \int d^d r \; \bigg(\frac{\tau}{2} -r_0 \bigg)^{j_1+j_2} \bigg(\bigg(\frac{\tau}{2} -r^0 \bigg)^2 - \vert \vec{r} \vert^2 \bigg)^{p_0/2}  \bigg) \eeq
Then, in place of the integral, we will put right hand side of Eq \ref{c=1}. Since the power inside the integral is $j$ rather than $p_0$, we will have to replace $p_0$ with $j$ in the Eq \ref{c=1}. At the same time, \emph{outside} that expression, $p_0$ will remain $p_0$. If we make sure to do that, we obtain the following expression:  
\beq \int_{\alpha (p,q)} d^d r \; \bigg(\bigg(\frac{\tau}{2} -r^0 \bigg)^2 - \vert \vec{r} \vert^2 \bigg)^{p/2} \bigg(\frac{c_1\tau}{2} -x^0 \bigg)^{p_{01}} \bigg(\frac{c_2 \tau}{2} -x^0 \bigg)^{p_{02}} (x^1)^{p_1} \cdots (x^{d-1})^{p_{d-1}} =  \nonumber \eeq
\beq =  \sum_{j_1=1}^{p_{01}} \sum_{j_2=1}^{p_{02}}\bigg( {p_{01} \choose j_1} {p_{02} \choose j_2} (c_1-1)^{p_{01}-j_1}(c_2-1)^{p_{02}-j_2} \bigg(\frac{\tau}{2} \bigg)^{p_{01}+p_{02}-j_1-j_2} \times \nonumber \eeq
\beq \times \frac{ (-1)^{d-2+p_1+ \cdots + p_{d-1}} S_{p_1 \cdots p_{d-1}}\tau^{d+p+j_1+j_2 +p_1+ \cdots + p_{d-1}}}{2^{d-1+j_1+j_2+p_1+\cdots+p_{d-1}}}  \times  \nonumber \eeq
\beq \times  \sum_{k=0}^{d-2+p_1+ \cdots + p_{d-1}} \; \; \; \sum_{l=0}^{j_1+j_2+1} \bigg( {d-2+p_1+ \cdots + p_{d-1} \choose k} {j_1+j_2+1 \choose l} \times \nonumber \eeq
 \beq \times \frac{ \frac{1}{p-k+1}  - \frac{1}{d+p+j_1+j_2+ p_1 + \cdots + p_{d-1}}}{(k+d-1+j_1+j_2+p_1+ \cdots + p_{d-1}} \bigg)  \bigg] \nonumber \eeq
If we now combine $\tau$-s and $2$-s via the formula 
\beq \tau^{p_{01}+p_{02}-j_1-j_2} \tau^{d+p+j_1+j_2 +p_1 + \cdots + p_{d-1}} = \tau^{d+p+p_{01}+p_{02} + p_1+ \cdots + p_{d-1}} \eeq 
\beq 2^{p_{01}+p_{02}-j_1-j_2} 2^{d-1+j_1+j_2+p_1+ \cdots +p_{d-1}} = 2^{d-1+p_{01}+p_{02}+p_1+ \cdots + p_{d-1}} \eeq 
we obtain 
\beq \int_{\alpha (p,q)} d^d r \; \bigg(\bigg(\frac{\tau}{2} -r^0 \bigg)^2 - \vert \vec{r} \vert^2 \bigg)^{p/2} \bigg(\frac{c_1\tau}{2} -x^0 \bigg)^{p_{01}} \bigg(\frac{c_2\tau}{2} -x^0 \bigg)^{p_{02}}(x^1)^{p_1} \cdots (x^{d-1})^{p_{d-1}} =  \nonumber \eeq
\beq =  \sum_{j_1=1}^{p_{01}} \sum_{j_2=1}^{p_{02}} \bigg[ {p_{01} \choose j_1}{p_{02} \choose j_2} (c_1-1)^{p_{01}-j_1}(c_2-1)^{p_{02}-j_2} \times \nonumber \eeq
\beq \times \frac{ (-1)^{d-2+p_1+ \cdots + p_{d-1}} S_{p_1 \cdots p_{d-1}}\tau^{d+p+p_{01}+p_{02} +p_1+ \cdots + p_{d-1}}}{2^{d-1+p_{01}+p_{02}+p_1+ \cdots + p_{d-1}}}  \times  \nonumber \eeq
\beq \times  \sum_{k=0}^{d-2+p_1+ \cdots + p_{d-1}} \; \; \; \sum_{l=0}^{j_1+j_2+1} \bigg( {d-2+p_1+ \cdots + p_{d-1} \choose k} {j_1+j_2+1 \choose l} \times \nonumber \eeq
\beq \times  \frac{ \frac{1}{p-k+1}  - \frac{1}{d+p+j_1+j_2+ p_1 + \cdots + p_{d-1}}}{(k+d-1+j_1+j_2+p_1+ \cdots + p_{d-1}} \bigg)  \bigg]=  \nonumber \eeq
\beq =  \frac{(-1)^{d-2+p_1+ \cdots + p_{d-1}} S_{p_1 \cdots p_{d-1}}\tau^{d+p+p_0 + \cdots + p_{d-1}}}{2^{d-1+p_0+ \cdots + p_{d-1}}} \sum_{j=1}^{p_0} \bigg[  {p_{01} \choose j_1} {p_{02} \choose j_2}  (c_1-1)^{p_{01}-j}  (c_2-1)^{p_{02}-j}  \times  \nonumber \eeq
\beq \times  \sum_{k=0}^{d-2+p_1+ \cdots + p_{d-1}} \; \; \; \sum_{l=0}^{j+1} \bigg( {d-2+p_1+ \cdots + p_{d-1} \choose k} {j_1+j_2+1 \choose l} \frac{ \frac{1}{p-k+1}  - \frac{1}{d+p+j+ p_1 + \cdots + p_{d-1}}}{(k+d-1+j+p_1+ \cdots + p_{d-1}} \bigg)  \bigg]  \nonumber = \label{IntegralComplicatedFinal} \eeq
\beq = T_{d; c_1,c_2; p; p_{01},p_{02};p_1, \cdots, p_{d-1}} \tau^{d+p+p_{01} +p_{02}+p_1+ \cdots + p_{d-1}} \label{ExtractFromT} \eeq
where 
\beq  T_{d; c_1,c_2; p; p_{01},p_{02};p_1, \cdots, p_{d-1}} =  \frac{(-1)^{d-2+p_1+ \cdots + p_{d-1}} S_{p_1 \cdots p_{d-1}}}{2^{d-1+p_0+ \cdots + p_{d-1}}} \sum_{j=1}^{p_0} \bigg[  {p_{01} \choose j_1} {p_{02} \choose j_2}  (c_1-1)^{p_{01}-j}  (c_2-1)^{p_{02}-j}  \times  \nonumber \eeq
\beq \times  \sum_{k=0}^{d-2+p_1+ \cdots + p_{d-1}} \; \; \; \sum_{l=0}^{j+1} \bigg( {d-2+p_1+ \cdots + p_{d-1} \choose k} {j_1+j_2+1 \choose l} \frac{ \frac{1}{p-k+1}  - \frac{1}{d+p+j+ p_1 + \cdots + p_{d-1}}}{(k+d-1+j+p_1+ \cdots + p_{d-1}} \bigg)  \bigg]   \label{IntegralComplicatedFinal} \eeq
where $S_{p_1 \cdots p_{d-1}}$ is given by Eq \ref{SFinal}.

\subsection*{Computing A/B}

We are now ready to go back to our original question and find out what should be the ratio of $A/B$ in order for Lagrangian to be proportional to $F^{\mu \nu} F_{\mu \nu}$. Let us now substitute Eq \ref{IntegralComplicatedFinal} into Eq \ref{FewerTerms}

\beq  \int_{p \prec r \prec s \prec q} d^d r d^d s \; \Phi^2_{prsqp} = \nonumber \eeq
\beq = \frac{\tau^{2d+4}}{2} F_{0k} F_{0k} \bigg[ I_{d2} T_{d;0,1;d;2,2; 0, \cdots, 0}  + \bigg(\frac{I_{d0}}{2} + I_{d2}+2J_{d2} \bigg) T_{d;0,0;d;0,2;2,0, \cdots, 0}+ \frac{3}{4} I_{d0}   T_{d;0,0;d;0,0;2,0, \cdots,0}  \bigg] + \nonumber \eeq 
\beq + \frac{\tau^{2d+4}}{4} F_{ij} F_{ij} \bigg[ I_{d2} T_{d;0,0; d+2; 0,0; 2, 0, \cdots,0}  - \bigg( J_{d2}+ \frac{I_{d0}}{4} \bigg) T_{d;0,0; d; 0,0; 2,2,0, \cdots,0}\bigg]  \eeq
On the other hand, Eq \ref{ElectricSquare} tells us 
\beq  \int_{p \prec r \prec s \prec q} d^d r d^d s \; \Phi^2_{prqsp}  = \frac{I_{d0} I_{d2}}{2}  \tau^{2d+4} F_{0k} F_{0k} \eeq
Therefore, 
\beq \frac{A}{B} \int_{p \prec r \prec q \succ s \succ p} d^d r d^d s \; \Phi_{prqsp} +  \int_{p \prec r \prec s \prec q  \succ p} d^d r d^d s \; \phi_{prsqp} = \nonumber \eeq
\beq = \frac{\tau^{2d+4}}{2} F_{0k} F_{0k} \bigg[ \frac{A}{B} I_{d0} I_{d2} + I_{d2} T_{d;0,1;d;2,2; 0, \cdots, 0}  + \nonumber \eeq
\beq + \bigg(\frac{I_{d0}}{2} + I_{d2}+2J_{d2} \bigg) T_{d;0,0;d;0,2;2,0, \cdots, 0}+ \frac{3}{4} I_{d0}  T_{d;0,0;d;0,0;2,0, \cdots,0}  \bigg] + \nonumber \eeq 
\beq + \frac{\tau^{2d+4}}{4} F_{ij} F_{ij} \bigg[  I_{d2} T_{d;0,0; d+2; 0,0; 2, 0, \cdots,0}  - \bigg( J_{d2}+ \frac{I_{d0}}{4} \bigg) T_{d;0,0; d; 0,0; 2,2,0, \cdots,0}  \bigg]  \eeq
In order to have the above expression Lorentz covariant, we need to have 
\beq \frac{A}{B} I_{d0} I_{d2} + I_{d2} T_{d;0,1;d;2,2; 0, \cdots, 0}   + \bigg(\frac{I_{d0}}{2} + I_{d2}+2J_{d2} \bigg) T_{d;0,0;d;0,2;2,0, \cdots, 0}+ \frac{3}{4}I_{d0}  T_{d;0,0;d;0,0;2,0, \cdots,0} = \nonumber \eeq
\beq = - \frac{1}{2} \bigg( I_{d2} T_{d;0,0; d+2; 0,0; 2, 0, \cdots,0}  - \bigg( J_{d2}+ \frac{I_{d0}}{4} \bigg) T_{d;0,0; d; 0,0; 2,2,0, \cdots,0}  \bigg) \eeq 
By dividing this equation by $I_{d0} I_{d2}$, we obtain 
\beq \frac{A}{B} + \frac{T_{d;0,1;d;2,2; 0, \cdots, 0}}{I_{d0}}  + \bigg(\frac{1}{2I_{d2}} + \frac{1}{I_{d0}}+\frac{2J_{d2}}{I_{d0} I_{d2}} \bigg) T_{d;0,0;d;0,2;2,0, \cdots, 0}+ \frac{3}{4I_{d2}} T_{d;0,0;d;0,0;2,0, \cdots,0} = \nonumber \eeq
\beq = - \frac{1}{2} \bigg( \frac{1}{I_{d0}} T_{d;0,0; d+2; 0,0; 2, 0, \cdots,0}  - \bigg( \frac{J_{d2}}{I_{d0}I_{d2}}+ \frac{1}{4I_{d2}} \bigg) T_{d;0,0; d; 0,0; 2,2,0, \cdots,0}  \bigg) \eeq 
By keeping $A/B$ on the left hand side and moving everything else on the right hand side, we obtain
\beq \frac{A}{B} = - \frac{T_{d;0,1;d;2,2; 0, \cdots, 0}}{I_{d0}}  - \bigg(\frac{1}{2I_{d2}} + \frac{1}{I_{d0}}+\frac{2J_{d2}}{I_{d0} I_{d2}} \bigg) T_{d;0,0;d;0,2;2,0, \cdots, 0}- \frac{3}{4I_{d2}} T_{d;0,0;d;0,0;2,0, \cdots,0} - \nonumber \eeq
\beq - \frac{1}{2 I_{d0}} T_{d;0,0; d+2; 0,0; 2, 0, \cdots,0}  + \frac{1}{2} \bigg( \frac{J_{d2}}{I_{d0}I_{d2}}+ \frac{1}{4I_{d2}} \bigg) T_{d;0,0; d; 0,0; 2,2,0, \cdots,0}  \label{AlmostDoneAB} \eeq 
where we got rid of extra brackets by multiplying things term by term by what used to be an overall coefficient $-1/2$. By comparing Eq \ref{Setup1}, \ref{Setup2} and \ref{Setup3} to Eq \ref{ExtractFromT}, we obtain 
\beq I_{d0} = T_{d; 0,0; 0;0,0; 0, \cdots, 0} \label{IFromT1}\eeq
\beq I_{d2} = T_{d; 0,0; 0;0,0;2, 0, \cdots, 0} \label{IFromT2} \eeq
\beq J_{d2} = T_{d; 0,0; 0;2,0;0, \cdots, 0} \label{IFromT3} \eeq
Substituting Eq \ref{SFinal} into Eq \ref{IntegralComplicatedFinal}, then  substituting Eq \ref{IntegralComplicatedFinal} into $T$-s, and substituting Eq \ref{IFromT1}, \ref{IFromT2} and \ref{IFromT3} into Eq \ref{AlmostDoneAB} would produce the value of $A/B$. 

\subsection*{Scalar Field}

As was discussed in second and third section of this paper, the fact that we were using edges rather than points allowed us to address some of the locality and linearity issues. For this reason we suggest that edges should be used for scalar field as well. Let me briefly demonstrate how this can be done. Given points $r \prec s$ connected by an edge, I will define $\phi (r,s)$ as 
\beq \phi (r, s) = \frac{\int_{\gamma (r,s)} \phi (\gamma (\tau)) d \tau}{\int_{\gamma (r,s)} d \tau} \eeq
where $\gamma (r,s)$ is a geodesic connecting $r$ and $s$. Therefore, if $\phi$ is linear, we have 
\beq \phi \; {\rm is \; linear} \; \Longrightarrow \; \phi (r,s) = \phi \bigg(\frac{r+s}{2} \bigg) \eeq 
where on the left hand side $\phi$ is a function of an edge, while on the right hand side it is a function of a point. Thus, there is one to one correspondence between a point and an edge. However, the probability distribution of finding the resulting points will no longer be Poisson, but, instead, it will go up towards the middle of Alexandrov set and down towards its boundary. After all, if we take a point in the middle, there will be more ways of representing it as a midpoint of a segment, than if we take a point near the boundary. This feature is useless, and in fact it only complicates calculations. So the \emph{only} intention of what I am trying to do is that I make sure I think of those things as edges rather than points and \emph{then} I can use the fact that they are edges to address locality issues. Once again, we have an Alexandrov set $\alpha (p,q)$ where, as usual, we pick coordinate system in such a way that 
\beq p = \bigg(- \frac{\tau}{2}, 0, \cdots, 0 \bigg) \; , \; q = \bigg(\frac{\tau}{2}, 0, \cdots, 0 \bigg) \eeq 
Now, for a point $r$ inside the Alexandrov set, we have 
\beq \phi (p, r) = \phi \bigg(\frac{p+r}{2} \bigg) = \phi (0) + \frac{p^0+r^0}{2} \partial_0 \phi + \frac{p^k + r^k}{2} \partial_k \phi = \nonumber \eeq 
\beq = \phi (0) + \frac{- \frac{\tau}{2} + r^0}{2} \partial_0 \phi + \frac{0+r^k}{2} \partial_k \phi = \phi (0) - \frac{\tau}{4} \partial_0 \phi + \frac{r^0}{2} \partial_0 \phi + \frac{r^k}{2} \partial_k \phi \eeq 
and, similarly, 
\beq \phi (r, q) = \phi \bigg(\frac{q+r}{2} \bigg) = \phi (0) + \frac{q^0+r^0}{2} \partial_0 \phi + \frac{q^k + r^k}{2} \partial_k \phi = \nonumber \eeq 
\beq = \phi (0) + \frac{ \frac{\tau}{2} + r^0}{2} \partial_0 \phi + \frac{0+r^k}{2} \partial_k \phi = \phi (0) + \frac{\tau}{4} \partial_0 \phi + \frac{r^0}{2} \partial_0 \phi + \frac{r^k}{2} \partial_k \phi \eeq 
and, finally, 
\beq \phi (p,q) = \phi (0) \eeq 
From these we find
\beq \phi (r,q)- \phi (p,r) = (1-1) \phi (0) + (1+1) \frac{\tau}{4} \partial_0 \phi + (1-1) \frac{r^0}{2} \partial_0 \phi + (1-1) \frac{r^k}{2} \partial_k \phi = \frac{\tau}{2} \partial_0 \phi \label{TimelikeSeparatedEdges} \eeq 
and also 
\beq \frac{\phi (p, r) + \phi (r,q)}{2} - \phi (p,q) = \nonumber \eeq
 \beq = \frac{(1+1) \phi (0) + (1-1) \frac{\tau}{4} \partial_0 \phi + (1+1) \frac{r^0}{2} \partial_0 \phi + (1+1) \frac{r^k}{2} \partial_k \phi}{2} - \phi (0) = \nonumber \eeq
\beq = \phi_0 + \frac{r^0}{2} \partial_0 \phi + \frac{r^k}{2} \partial_k \phi - \phi (0) = \frac{r^0}{2} \partial_0 \phi + \frac{r^k}{2} \partial_k \phi \label{SpacelikeSeparatedEdges} \eeq 
If we integrate Eq \ref{TimelikeSeparatedEdges} we obtain
\beq \int d^d r \; (\phi (r,q) - \phi (p,r))^2 = \int_{\alpha (p,q)} d^d r \; \bigg(\frac{\tau}{2} \partial_0 \phi \bigg)^2 = \nonumber \eeq
\beq = \bigg(\frac{\tau}{2} \partial_0 \phi \bigg)^2 \int_{\alpha (p,q)} d^d r =\bigg(\frac{\tau}{2} \partial_0 \phi \bigg)^2 I_{d0} \tau^d = \frac{I_{d0}}{4} \tau^{d+2} (\partial_0 \phi)^2 \label{ScalarTimelike} \eeq 
and if we integrate Eq \ref{SpacelikeSeparatedEdges} we obtain
\beq \int_{\alpha (p,q)} d^d r \; \bigg(\frac{\phi (p, r) + \phi (r,q)}{2} - \phi (p,q)  \bigg)^2 = \int_{\alpha (p,q)} d^d r \; \bigg( \frac{r^0}{2} \partial_0 \phi + \frac{r^k}{2} \partial_k \phi \bigg)^2 = \nonumber \eeq 
\beq = \int_{\alpha (p,q)} d^d r \; \bigg( \frac{1}{4} (\partial_0 \phi)^2 (r^0)^2 + \frac{1}{2} (\partial_0 \phi) (\partial_k \phi) r^0r^k + \frac{1}{4} (\partial_k \phi)^2 (r^k)^2 \bigg) \eeq 
By symmetry consideration, $r^0 r^k$ integrates to zero, so we obtain 
\beq \int_{\alpha (p,q)} d^d r \; \bigg(\frac{\phi (p, r) + \phi (r,q)}{2} - \phi (p,q)  \bigg)^2 =  \nonumber \eeq
 \beq = \int_{\alpha (p,q)} d^d r \; \bigg( \frac{1}{4} (\partial_0 \phi)^2 (r^0)^2 + \frac{1}{4} (\partial_k \phi)^2 (r^k)^2 \bigg) = \nonumber \eeq 
\beq =\frac{1}{4} (\partial_0 \phi)^2 \int_{\alpha (p,q)} d^d r \; (r^0)^2 + \frac{1}{4} (\partial_k \phi)^2 \int d^d r  = \nonumber \eeq 
\beq = \bigg( \frac{1}{4} (\partial_0 \phi)^2 \bigg) I_{d0} \tau^{d+2} + \bigg(\frac{1}{4} (\partial_k \phi)^2 \bigg) I_{d2} \tau^{d+2} = \nonumber \eeq 
\beq = \frac{\tau^{d+2}}{4} \big(I_{d0} (\partial_0 \phi)^2 + I_{d2} (\partial_k \phi)^2 \big) \label{ScalarSpacelike} \eeq
And then we find the appropriate linear combination of Eq \ref{ScalarTimelike} and \ref{ScalarSpacelike} to get $\partial^{\mu} \phi \partial_{\mu} \phi$. 

\subsection*{Conclusion}

In this paper we have proposed the Lagrangian of scalar and gauge fields on causal sets. While similar work has been done in \cite{GravityAndMatter},  \cite{Gauge} and \cite{Bosonic}, the difference between the current paper and those other papers is that we have restricted ourselves to timelike two-point functions: that is, $a (r,s)$ is only defined if either $r \prec s$ or $s \prec r$ (in contrast to \cite{Gauge} and \cite{Bosonic} where $a(r,s)$ is defined for all pairs $(r,s)$) and the ``scalar" field $\phi$ is viewed as a two-point function as well (in contrast to \cite{GravityAndMatter} and \cite{Bosonic} where $\phi (x)$ is a single point function). Apart from that, the assumption about the number of points contained in $\alpha (p,q)$ is also different. As far as \cite{GravityAndMatter}, \cite{Gauge} and \cite{Bosonic} are concerned, $\alpha (p,q)$ is small enough for all fields to be linear, yet large enough to contain a very large number of points so that the spacetime region ``looks like a continuum". On the other hand, in the current paper, the region $\alpha (p,q)$ typically contains only a couple of points other than $p$ and $q$ itself: the statistical role of large number of points is replaced by the fact that there are large number of choices of $p$ and $q$ within the larger region of spacetime we happened to be concerned about. In some sense this makes the theory look more physical, since the interaction that the Lagrangian implies is now more direct, and also is along causal lines. Apart from that, in case of spacelike edge the set of Lorentz transformations in the hyperplane perpendicular to the edge isn't compact, while in case of timelike edge it is, and this makes it easier to resolve non-locality issues if we limit ourselves to just the timelike edges. 

This brings us to the other implication of the proposal: a way of addressing nonlocality issues. As discussed in \cite{Nonlocality}, there is a considerable difficulty of combining discreteness, locality and Lorentz invariance. In this paper we found a way to ``have our cake and eat it too": on the one hand, we have violated relativity, and thus restored locality, by focusing on the edge rather than a point. But, on the other hand, we can argue that we didn't really violate relativity since the edges are present in any causal set, and they always play some role (for example, they are needed in defining the holonomy $a(r,s)$); we simply made them have bigger role in this paper by putting $\phi$ and $\cal L$ into the same category as $a$. A continuum version of this argument is that we replace the Lagrangian density ${\cal L} (\phi, A^{\mu}; x^{\mu})$ with ${\cal L} (\phi, A^{\mu}; x^{\mu}, v^{\mu})$, where $v^{\mu}$ is a tangent vector at $x^{\mu}$. The relation between discrete and continuum is given by 
\beq \frac{p^{\mu} + q^{\mu}}{2} \longrightarrow x^{\mu} \eeq
\beq \frac{q^{\mu}- p^{\mu}}{\tau (p,q)} \longrightarrow v^{\mu} \eeq
Now, when we have ${\cal L} (A^{\mu}, \phi; x^{\mu})$, we aren't worried that $x^{\mu}$ would violate translational invariance. Analogously, we shouldn't be concerned that $v^{\mu}$ in ${\cal L} (\phi, A^{\mu}; x^{\mu}, v^{\mu})$ should violate Lorentz invariance either. Indeed, this argument has also been used in \cite{PhaseSpacetime} but in a different way: in that paper the points themselves were associated with $(x^{\mu}, v^{\mu})$ rather than $x^{\mu}$ since the Poisson scattering over a manifold was replaced with a Poisson scattering over a tangent bundle. Thus, in some vague sense, a point in \cite{PhaseSpacetime} plays the same role as the edge in this paper. Other than this vague relation, the details of the theories are different. 

\subsection*{Achnowledgements}

The author is greateful to Professor Luca Bombelli (University of Mississippi) and Professor Vassilev (University of New Mexico) for helpful discussion. The research of Roman Sverdlov was supported in part by NSF grant DMS-1554456 and DMS 9215024


\begin{thebibliography}{77}

\bibitem{Review1} L Bombelli, J Lee, D Meyer and R Sorkin 1987 “Space-time as a causal set” Phys. Rev.
Lett. 59 521-524.

\bibitem{Review2} D.D. Reid; Introduction to causal sets: an alternate view of spacetime structure; Canadian Journal of Physics 79, 1-16 (2001); arXiv:gr-qc/9909075; (General)

\bibitem{Review3} F. Dowker, Introduction to causal sets and their phenomenology, Gen Relativ Gravit (2013) 45:1651–1667 doi:10.1007/s10714-013-1569-y

\bibitem{Review4} J. Henson, The causal set approach to quantum gravity, arXiv:gr-qc/0601121

\bibitem{Hawking} S W Hawking, A R King and P J McCarthy 1976 “A new topology for curved spacetime
which incorporates the causal, differential and conformal structures” J. Math. Phys. 17 174-181.

\bibitem{Malament} D Malament 1977 “The class of continuous timelike curves determines the topology of
spacetime” J. Math. Phys. 18 1399-1404.

\bibitem{Dambertian1} R. Sorkin ``Scalar Field Theory on a Causal Set in Histories Form" Journal of Physics: Conference Series, Volume 306, Number 1

\bibitem{Dambertian2} F. Dawker, L. Glaser ``Causal set d'Alembertians for various dimensions"  J. Class. Quant. Grav. (2013) IOP Publishing Ltd
Classical and Quantum Gravity, Volume 30, Number 19

\bibitem{Nonlocality} Rafael Sorkin ``Does locality fail at intermediate length-scales?" arXiv:grqc/0703099

\bibitem{NonlocalityTheorem} L. Bombelli, J. Henson and R. Sorkin, “Discreteness without Symmetry Breaking: a
Theorem,” arXiv:gr-gc/060500v1

\bibitem{GravityAndMatter} R Sverdlov and L Bombelli, "Gravity and matter in causal set theory," arXiv:0801.0240 (2008), and Class. Quantum Grav. 26: 075011 (2009).

\bibitem{Gauge} Roman Sverdlov, ``Gauge Fields in Causal Set Theory" arXiv:0807.2066

\bibitem{Bosonic} Roman Sverdlov, ``Introduction of Bosonic Fields into Causal Set Theory" Workshop on Continuous and Lattice Approaches to Quantum Gravity 17-19 September 2008 $http://pos.sissa.it/archive/conferences/079/014/CLAQG08_014.pdf$ 2008 and  arXiv:0807.4709

\bibitem{Smolin1} M. Cortes, L. Smolin ``Energetic Causal Sets" Phys. Rev. D 90, 044035 (2014) and arXiv:1308.2206

\bibitem{Smolin2} M. Cortes, L. Smolin ``Quantum Energetic Causal Sets" Phys. Rev. D 90, 044035 – Published 14 August 2014

\bibitem{Johnston1} S. Johnston, ``Particle propagators on discrete spacetime" Class.Quant.Grav.25:202001,2008

\bibitem{Johnston2} S. Johnston, ``Quantum fields on causal sets" PhD Thesis, Imperial College London, September 2010 and arXiv:1010.5514

\bibitem{PhaseSpacetime} Roman Sverdlov, ``Causal set as discretized phase spacetime" arXiv:0910.2498


\end{thebibliography}
\end{document}